\documentclass[%
preprint,
 amsmath,amssymb,
 aps,
]{revtex4-2}

\usepackage{epstopdf} 

\usepackage{graphicx}

\usepackage{dcolumn}
\usepackage{bm}
\usepackage{hyperref}

\usepackage{epstopdf}
\usepackage{lineno}
\usepackage{textcomp}
\usepackage[gen]{eurosym}
\usepackage{slashed}
\usepackage[dvipsnames]{xcolor}  
\usepackage{placeins}

\usepackage{fancyheadings}
\usepackage{lastpage}
\usepackage{caption} 
\usepackage{verbatim}
\usepackage{color} 
\usepackage{subcaption} 
\usepackage{datetime}
\usepackage{booktabs, tabularx, multirow, array} 
\usepackage[separate-uncertainty,retain-explicit-plus,per-mode=symbol,binary-units]{siunitx}
\usepackage[version=4]{mhchem}
\DeclareSIUnit\c{\mbox{$c$}}
\DeclareSIUnit\year{yr}

\begin{document}

\preprint{APS/123-QED}

\title{Characterization of FBK NUV-HD-Cryo SiPMs near LHe temperature}
\author{Fengbo Gu}
\email{FengboGu@outlook.com}
\affiliation{%
Department of Nuclear Physics, China Institute of Atomic Energy, Sanqiang Rd. 1, Fangshan district, Beijing, China, 102413. 
}%

\author{Junhui Liao}
\email{junhui.private@gmail.com}
\affiliation{%
Department of Nuclear Physics, China Institute of Atomic Energy, Sanqiang Rd. 1, Fangshan district, Beijing, China, 102413. \\
Department of Physics, Brown University, Hope St. 182, Providence, Rhode Island, USA, 02912.\\
Yalong River Hydropower Development Company, Ltd., 288 Shuanglin Road, Chengdu 610051, China.\\
Jinping Deep Underground Frontier Science and Dark Matter Key Laboratory of Sichuan Province, Liangshan 615000, China.
}%

\author{Meiyunan Ma, Zhuo Liang, Zhaohua Peng, Jian Zheng, Jiangfeng Zhou, Guangpeng An}
\affiliation{%
Department of Nuclear Physics , China Institute of Atomic Energy, Sanqiang Rd. 1, Fangshan district, Beijing, China 
}%

\author{Lifeng Zhang, Lei Zhang}
\affiliation{%
Department of Nuclear Synthesis Technology, China Institute of Atomic Energy, Sanqiang Rd. 1, Fangshan district, Beijing, China 
}%
\author{Yuanning Gao}
\affiliation{%
 School of Physics, Peking University 
}%
\author{Fabio Acerbi, Andrea Ficorella, Alberto Gola, Laura Parellada Monreal}
\affiliation{%
Fondazione Bruno Kessler, via Sommarive, 18, Trento, 38123, Italy 
}%

\date{\today}

\begin{abstract}
Five FBK ``NUV-HD-Cryo'' SiPMs have been characterized at 7 K and 10 K, with 405 nm and 530 nm LED light, respectively. The dark count rate (DCR) was measured to be $\sim$ 1 Hz for the $\sim$ 100 mm$^2$-size SiPMs, or 0.01 Hz/mm$^2$, which is $\sim$ 7 orders lower than the DCR at room temperature (RT). Given the very low DCR at these cryogenic temperatures, we measured the SiPMs' I-V curves with such a method: illuminated the SiPMs with weak light, which differs from the conventional measurements at RT. Then, we measured the photo-detection efficiency (PDE), after-pulse (AP), and cross-talk (CT) with a bias voltage ranging from overvoltage (OV) 5 to 11 V. At the OV interval (5 to 11 V), the PDE was between 20\% - 45\%, and the AP and CT were both between $\sim$ 5\% and $\sim$ 20\%. With an OV higher than 10 V, the PDE would be $\ge$ 40\%, and the AP and CT are $\sim$ 20\%. Combining all of the measurements, we are confident that the SiPMs can be equipped as the photosensors on liquid helium detectors, including but not limited to the time projection chambers, which we have proposed in hunting for low-mass dark matter directly and beyond.
\end{abstract}

\maketitle


\section{Introduction} \label{sec1introduction}

Liquid helium (LHe) detectors have been proposed or employed to search for solar neutrinos~\cite{HERON1, HERON88, HERON96}, dark matter~\cite{GuoMckinsey13, Maris17, SpiceHeRald21, Biekert22, ALETHEIA-EPJP-2023, QUEST-DMC23}, and neutron electron dipole moment (nEDM)~\cite{Huffman2000, Ito2012, Ito2016, Phan20}. When a particle interaction takes place within LHe, it will emit scintillation light peaked at 80 nm~\cite{Kubota68, Benson18}. Given that no commercial photosensor is capable of detecting vacuum ultraviolet (VUV) light directly, a wavelength shifter such as tetraphenyl butadiene (TPB) is often implemented to convert it to visible light. As a rule of thumb, a photosensor should be as close as possible to the liquid or gas to avoid unnecessary light loss. Consequently, one of the requirements for the photosensor equipped on an LHe detector is that it should be functional near LHe temperature, $\sim$ 4.5 K~\footnote{For superfluid liquid helium detectors, the required temperature for a photosensor would be even lower than 4.5 K, and the exact temperature depends on the detector system.}. Previous measurements have shown that photomultipliers (PMTs) and silicon photomultipliers (SiPMs) can work at 4.5 K or below~\cite{Ito2012, Cardini14, Iwai19, ZhangSiPM2022}. However, to the best of our knowledge, there is no complete characterization of a photosensor near 4.5 K. Although PMTs have been widely equipped on liquid argon and liquid xenon time projection chambers (TPCs)~\cite{LZ2022, XENONnT-NR-2023, PandaX2021, DarkSide20k17, DEAP2019}, they are not suitable for LHe TPCs because helium gas can permeate through the glass into the vacuum space of the photosensor. If that happens, an increase in dark current and degradation of the breakdown voltage is foreseen; as a result, the PMTs will deteriorate or even out of work since a discharge could happen inside the tube~\cite{PMTsHamamatsu}. As such, SiPMs might be the only photosensor option for LHe TPCs.
 
In the manuscript, we present the characterization of Fondazione Brune Kessler (FBK) “NUV-HD-Cryo” SiPMs near LHe temperature, including a current-voltage (I-V) curve, the dark count rate (DCR), photo-detection efficiency (PDE), and the after-pulse (AP) and cross-talk (CT) probabilities. The paper will be organized as follows. We first explain the testing setup in section~\ref{secTestSetup}. The detailed tests of the I-V curve, DCR, PDE, AP, and CT at 10 K will be addressed in section~\ref{secCharacterization10K}. After finishing the measurements at 10 K, we updated our detector, which made the SiPMs reach a simulation-consistent lower temperature of 7 K. We re-tested most of the parameters at 7 K again, as shown in section~\ref{secCharacterization7K}. Our discussions and summaries are in section~\ref{summaryDiscussion}.
 
 \section{The testing setup introduction} \label{secTestSetup}

ALETHEIA, standing for A Liquid hElium Time projection cHambEr In dArk matter, aims to hunt for low-mass dark matter (100's MeV/c$^2$ to 10 GeV/c$^2$) with liquid helium-filled TPCs. To fully exploit the detector's potential, the equipped photosensors should have a PDE greater than 40\% near LHe temperature~\cite{ALETHEIA-EPJP-2023}. In addition, the I-V curve, DCR, AP, and CT must be measured to know whether the FBK NUV-HD-Cryo SiPMs' performance suits the ALETHEIA project (and other applications mentioned above). In previous works, the SiPMs made using the same technology have been fully tested at cryogenic temperatures, e.g. at liquid nitrogen temperature  (77 K) and down to 40 K~\cite{Acerbi17, Capasso20, Razeto22, Acerbi2023}. For most of the parameters we tested in this work, we have seen comparable results (or in line with the decreasing trends with reducing temperature).

\subsection{The setup} \label{secPDETestAt10KsubsecSetup}

\begin{figure}[!t]	 
	\centering
    \includegraphics[width=0.8\textwidth]{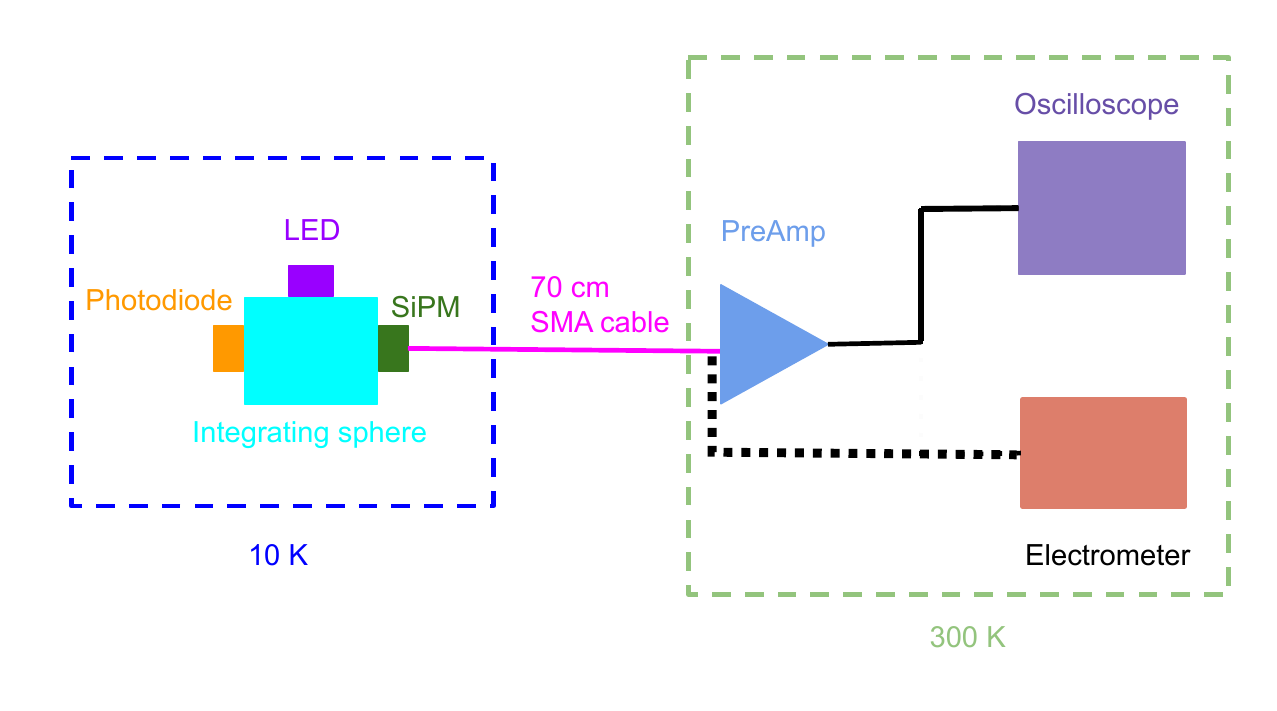}
	\caption{The schematic drawing of the test setup. The LED, the photodiode, and the SiPMs are mounted on the integrating sphere's surfaces, 90 degrees from each other. These parts are all at 10 K temperature. The 70 cm SMA cable conveys signals from cryogenic temperature to RT environment. Depending on the measurements, a preamplifier and a 8 GHz bandwidth, 25 GSa/s sample rate oscilloscope, or a Keithley 6485 electrometer will be connected for data-taking.}\label{schemePDESchematicDrawing} 
\end{figure}

As schematically shown in Fig.~\ref{schemePDESchematicDrawing}, the LED, the photodiode, and the SiPMs are all in a 10 K environment, while the preamplifier and data acquisition (DAQ) are at room temperature (RT). The 70 cm SMA cable going through the cryocooler’s walls connects the SiPMs in cryogenic conditions to the preamplifier on the warm side. All parts in cryogenic temperature have been verified earlier to ensure their capability of working at 4 K or below, and the parts were put in warm because they cannot work at LHe temperature. For the I-V curve tests, a Keithley 6485 electrometer was connected directly to the SiPMs; for the data-taking with the 8 GHz bandwidth, 25 GSa/s sample rate oscilloscope, an preamplifier was connected between the SiPMs and the oscilloscope.

Generally, a preamplifier should be located as close as possible to a detector to reduce signal deterioration on the cable. Reference~\cite{Acerbi17} used a preamplifier capable of working at 40 K, so the amplifier was (only) a few cm away from the SiPMs in cryogenic. High electron mobility transistors (HEMTs) have been implemented as a preamplifier at 4 K and 10 mK ~\cite{HEMTCryogenics19, HEMTCryogenics23}. In contrast, some other works~\cite{ZhangSiPM2022, Alessandro14} put the preamplifier at RT. In our setup, since the TI OPA694 chip integrated into the preamplifier cannot be functional at LHe temperature, we must leave the preamplifier at RT. We developed two versions of the preamplifier. The V1 amplifier is based on the ultra wide-band (690 MHz), low-power, current feedback operational amplifier, OPA694, and is configured as a non-inverting voltage amplifier. A low-pass RC filter, composed of a 20 $\Omega$ serial resistor and a 1 nF capacitor is added to the operational amplifier output to reduce noise further. The V2 preamplifier has been updated to have a lower bandwidth of $\sim$ 170 kHz, a greater gain, and a better signal/noise ratio (SNR). The V1 was used in the measurements of DCR (Fig.~\ref{SiPMDCR10k}) and AP (Fig.~\ref{SiPMAPDCR10k},  Fig.~\ref{SiPMAPVSBias10k}, and Fig.~\ref{combinedAP7K10K}). The V2 was implemented in the measurements of a raw signal (Fig.~\ref{typicalRawSignal10K}), a charge histogram (Fig.~\ref{SiPM10K35VQFit}), PDE (Fig.~\ref{allPDESiPM10K} and Fig.~\ref{combinedPDE7K10K}), and CT (Fig.~\ref{SiPMsCT7K10KOV5To11V}).

\begin{figure}[!t]	 
	\centering
    \includegraphics[width=0.8\textwidth]{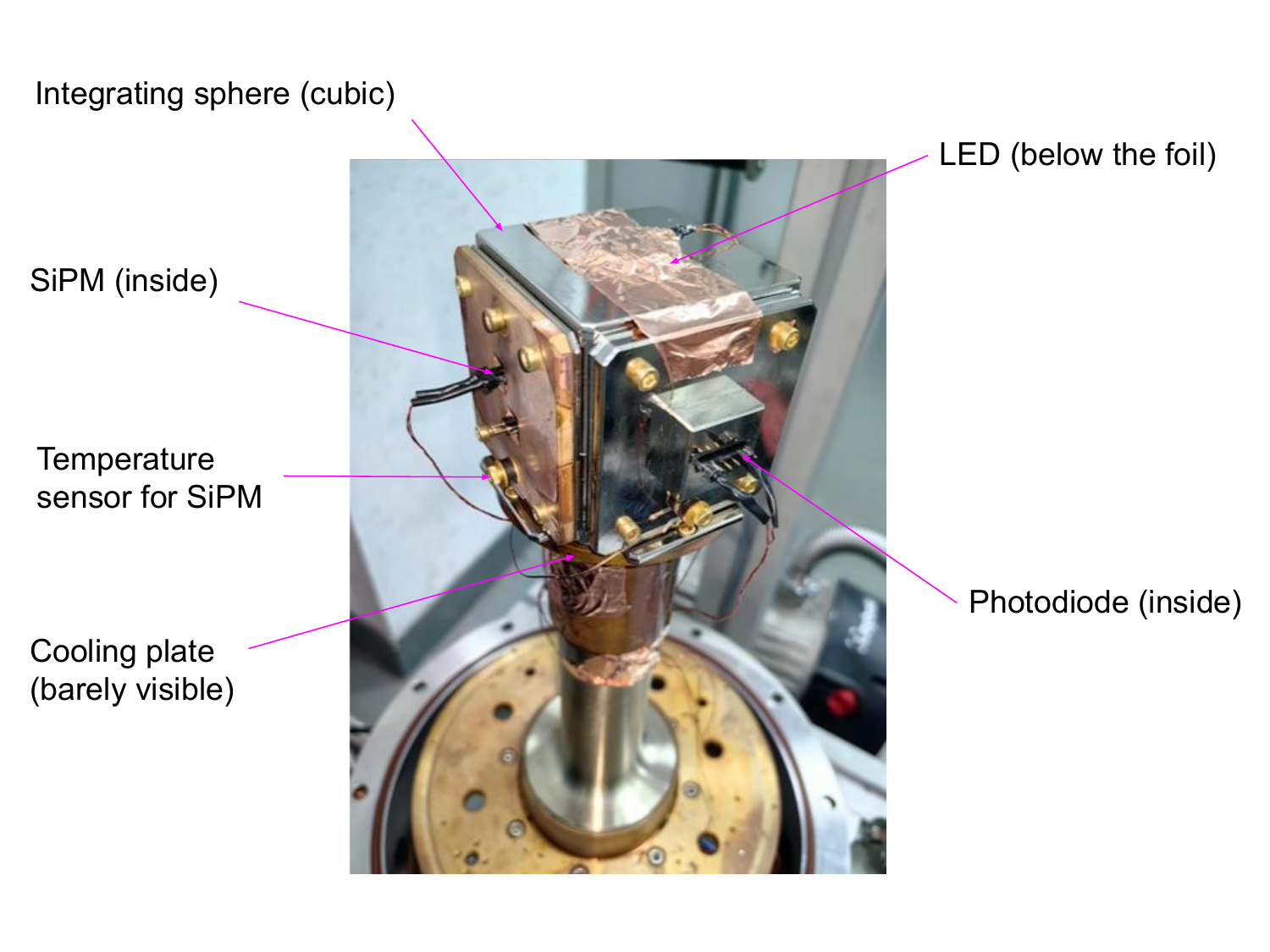}
	\caption{The picture shows the integrating sphere mounted on the cooling plate (inside the G-M cryocooler). The photodiode, the LED, and the SiPMs are also visible in the image.}\label{integratingSphereOnCoolingPlateRealPic} 
\end{figure}

Fig.~\ref{integratingSphereOnCoolingPlateRealPic} shows the actual setup; the cubic integrating sphere mounts on the barely visible cooling plate, which is the cooling source of the detector system. The LED, the photodiode, the SiPMs, and the temperature sensor (TS) are all installed on the surfaces of the sphere. Being driven by the pulses produced in a signal generator, DG535~\footnote{https://www.thinksrs.com/products/dg535.html}, the LEDs can emit monochromatic light with an intrinsic FWHM of $<$ 5\%. The NUV-HD-Cryo SiPMs, having an active area of 11.7 $\times$ 7.9 mm$^2$, a cell pitch of 30 $\times$ 30 $\mu$m$^2$, was thankfully provided by FBK. We bought the 10 $\times$ 10 mm$^2$ (active area) photodiode from Opto diode company~\footnote{https://optodiode.com}. For more info on the photodiode calibration, please refer to section~\ref{secPDETestAt10KsubsecPDCal}.

\begin{figure}[!t]	 
	\centering
    \includegraphics[width=0.8\textwidth]{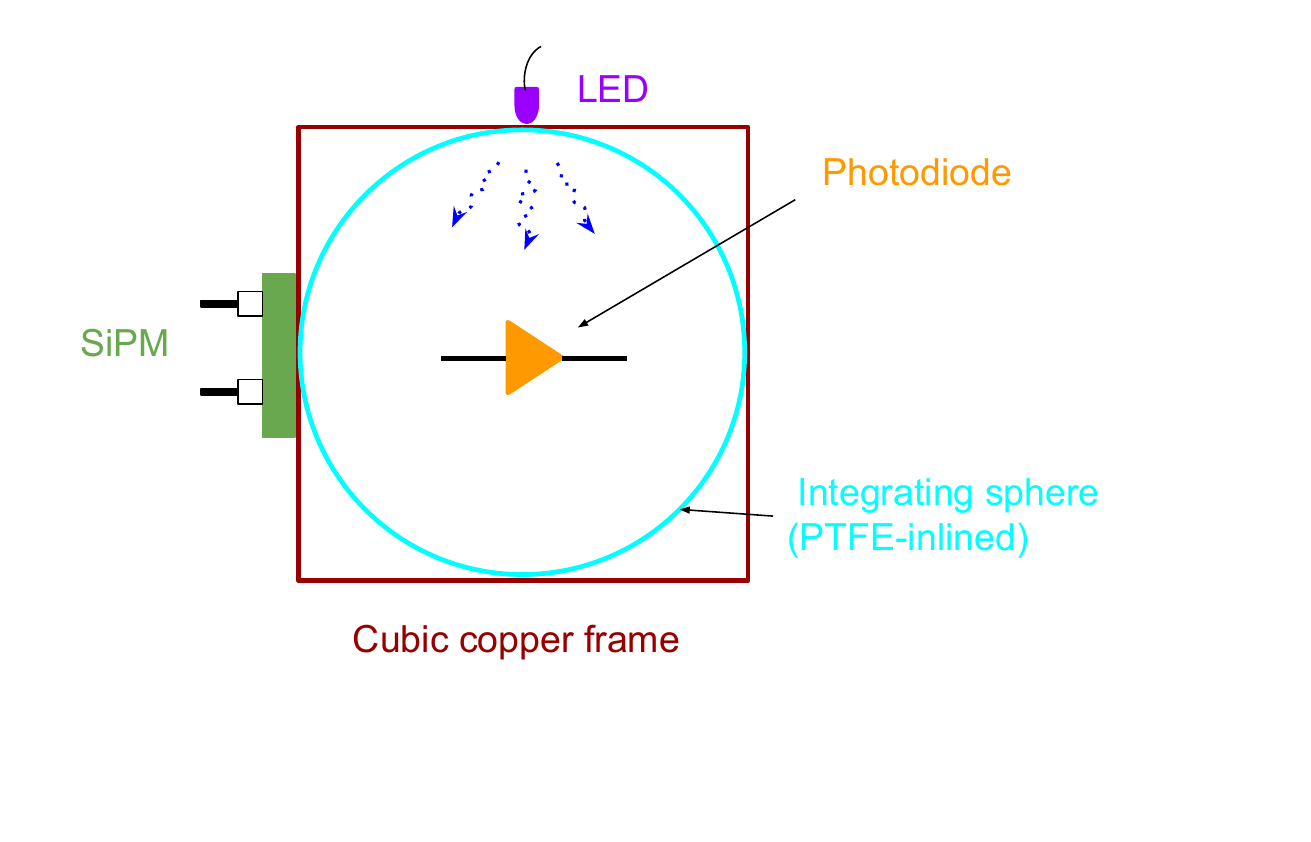}
	\caption{The schematic drawing shows the PDE test setup. The inner surface of the integrating sphere is PTFE (not visible in the drawing). The LED simultaneously illuminates the to-be-calibrated SiPMs and the calibrated photodiode. Please refer to the main text for more information.}\label{schemeIntegratingSphereSchematicDrawing} 
\end{figure}

Our team designed the integrating sphere and outsourced it to a commercial company in China. The inner surface of the 60 mm diameter integrating sphere is coated with highly reflective PTFE. TPB has been implemented as a wavelength shifter below 2 K in other works~\cite{SpiceHeRald21, Biekert22, Ito2012, Ito2016, Phan20}. The outside of the sphere is cubic, which gives convenience to (a) installing experimental parts on its surfaces and (b) being mounted securely on the cooling plate of the G-M cryocooler. 
As illustrated in Fig.~\ref{schemeIntegratingSphereSchematicDrawing}, the LED, the calibrated photodiode, and the to-be-calibrated SiPMs are positioned 90 degrees from each other (on different ports of the sphere). This design ensures that the photodiode and the SiPMs receive the fixed ratio of photons as long as the size of the two corresponding ports does not change, regardless of the intensity profile of light emitted from the LED. In our case, the diameters of the ports for the photodiode and the SiPMs are 4.0 mm and 0.1 mm, respectively.

The cooling plate of the G-M cryocooler approaches its lowest temperature, $\sim$ 3 K, after $\sim$ 3 hours of cooling; for the SiPMs on the integrating sphere, it will take a couple of additional hours to reach the lowest temperature. During our tests, we often run the cooling machine for more than 12 hours before launching a measurement to ensure the SiPMs and other parts on the sphere have reached the lowest temperature. Since it is impossible to measure the SiPMs' temperature accurately, in the paper, the SiPMs' temperature refers to the one measured on a TS $\sim$ 5 cm away. According to our simulation, the SiPMs' actual temperature would be $\sim$ 0.3 K higher than the sensor-measured value, as will be further discussed in section~\ref{secUnserstandSiPMsTemp}. We did not test the SiPMs at a temperature higher than 10 K because it is not the typical temperature the photosensor equipped on LHe detectors likely to work, though reference~\cite{Acerbi2023} tested the SiPMs between 75 K and 300 K; neither were the SiPMs characterized below 7 K due to the cryocooler's cooling power limitation.

\subsection{Photodiode responsivity calibration at NIM in China} \label{secPDETestAt10KsubsecPDCal}

As mentioned in references~\cite{Eckert10, Zappala16, MPPCTechnicalNote21}, a to-be-tested SiPMs' PDE can be figured out by measuring the relative efficiency between the SiPMs and a calibrated photodiode. That being said, the efficiency (responsivity) of the photodiode should be characterized in advance. The photodiode under test is AXUV100G from Opto diode company. Although the company's website has a datasheet, we cannot implement the responsivity data directly for a calibration near 4.5 K since the official data sheet was measured at RT. A typical option was to calibrate the photodiodes at the National Institute of Standards and Technology (NIST), but this did not happen because NIST do not offer such a calibration near LHe temperature for the moment, though they would be capable of doing that in the future~\cite{privateCommunicationsWithNIST}. Luckily, a team at the National Institute of Metrology (NIM) in China agreed to calibrate our photodiodes in their G-M cryocooler as a side-project of their primary research programs. Although the NIM facility can only calibrate a photodiode with 405 and 530 nm light, the calibration is still valuable since the two wavelengths line up well with TPB-shifted LHe scintillation, $\sim$ 350 - 600 nm~\cite{Benson18}.

\begin{figure}[!t]	 
	\centering
    \includegraphics[width=0.8\textwidth]{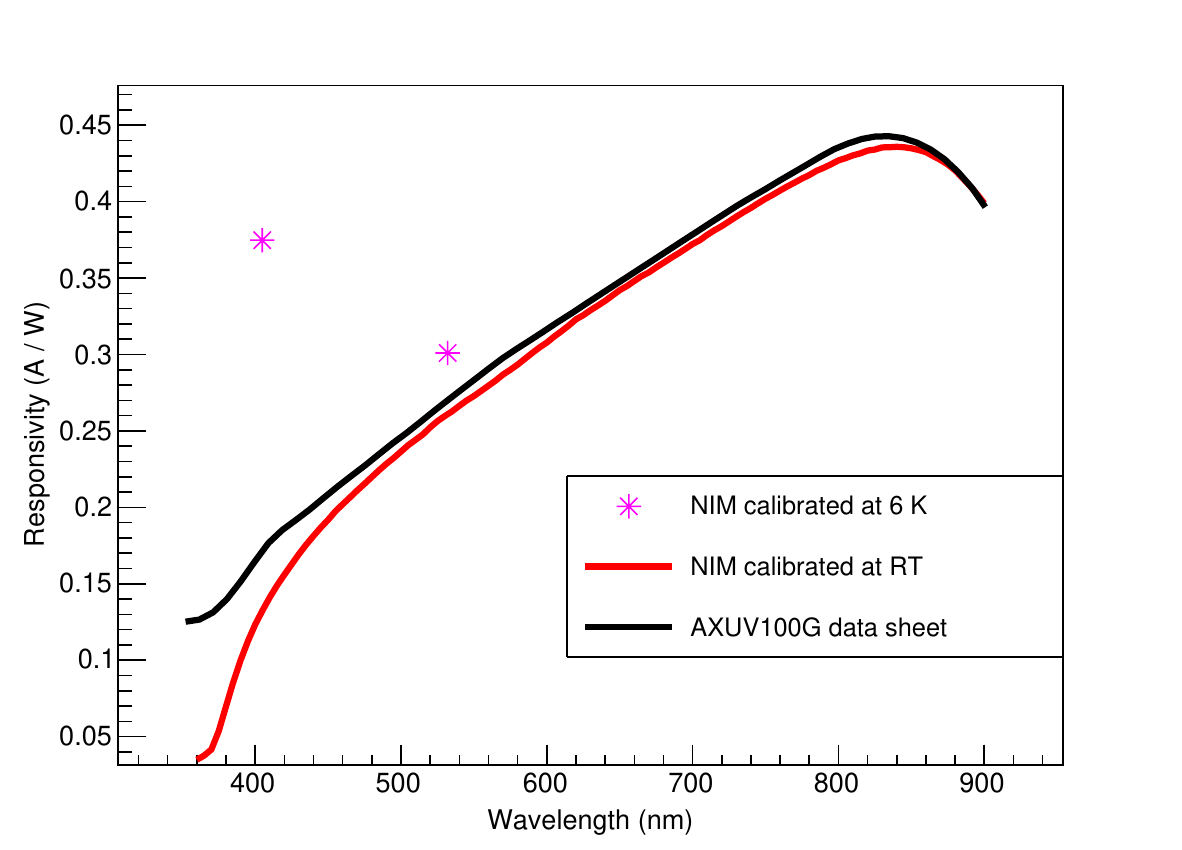}
	\caption{The responsivity of the photodiode we used for PDE measurements. The black curve is copied from the photodiode's datasheet. The red curve represents the data points measured at NIM at RT. The two stars were the photodiode's characterized responsivity when they were cooled to 6 K and illuminated with 405 nm and 530 nm light, respectively.}\label{figResponsivityCombined} 
\end{figure}

We first asked the NIM team to calibrate the responsivity at RT to see whether their calibration is consistent with the official datasheet. As shown in Fig.~\ref{figResponsivityCombined}, the red curve, representing the NIM calibrated results, overlaps well with the black curve, corresponding to the official datasheet. Then, two photodiodes were calibrated at 6 K, the lowest temperature the cryocooler can reach. The calibrated responsivity of the photodiode being implemented in the PDE tests as the two dots shown in Fig.~\ref{figResponsivityCombined}.

During the calibrations at NIM, the temperature sensor and the photodiode were mounted on the same copper frame and separated $\sim$ 8.0 cm. Our COMSOL simulation showed that the temperature difference between the photodiode and the temperature sensor was $\sim$ 0.1 K, which is negligible compared to the temperature sensor's 1.0 K uncertainty. Therefore, we consider the photodiode's actual temperature during the calibration was directly measured by the sensor, 6 K.

\section{Characterizing the SiPMs at 10 K} \label{secCharacterization10K}

We characterized the SiPMs' I-V curve, DCR, PDE, AP, and CT at 10 K in the winter of 2023. At that time, we thought 10 K was the lowest temperature the SiPMs could achieve until we launched a COMSOL simulation to understand the SiPMs' accurate temperature. The simulation showed the SiPMs should be around 7 K instead of 10 K. It turned out that a tiny gap existed on the integrating sphere's contacting surfaces, which induced cryogenics leakage. After filling the gaps with indium films, the SiPMs' temperature reached 7 K, which was consistent with the simulation. We re-measured the parameters at 7 K again. For more info, please refer to section~\ref{secCharacterization7K}.

\subsection{Testing the SiPMs' I-V curve at 10 K} \label{secPDETestAt10KsubsecIVCurve}

\begin{figure}[!t]	 
	\centering
    \includegraphics[width=0.8\textwidth]{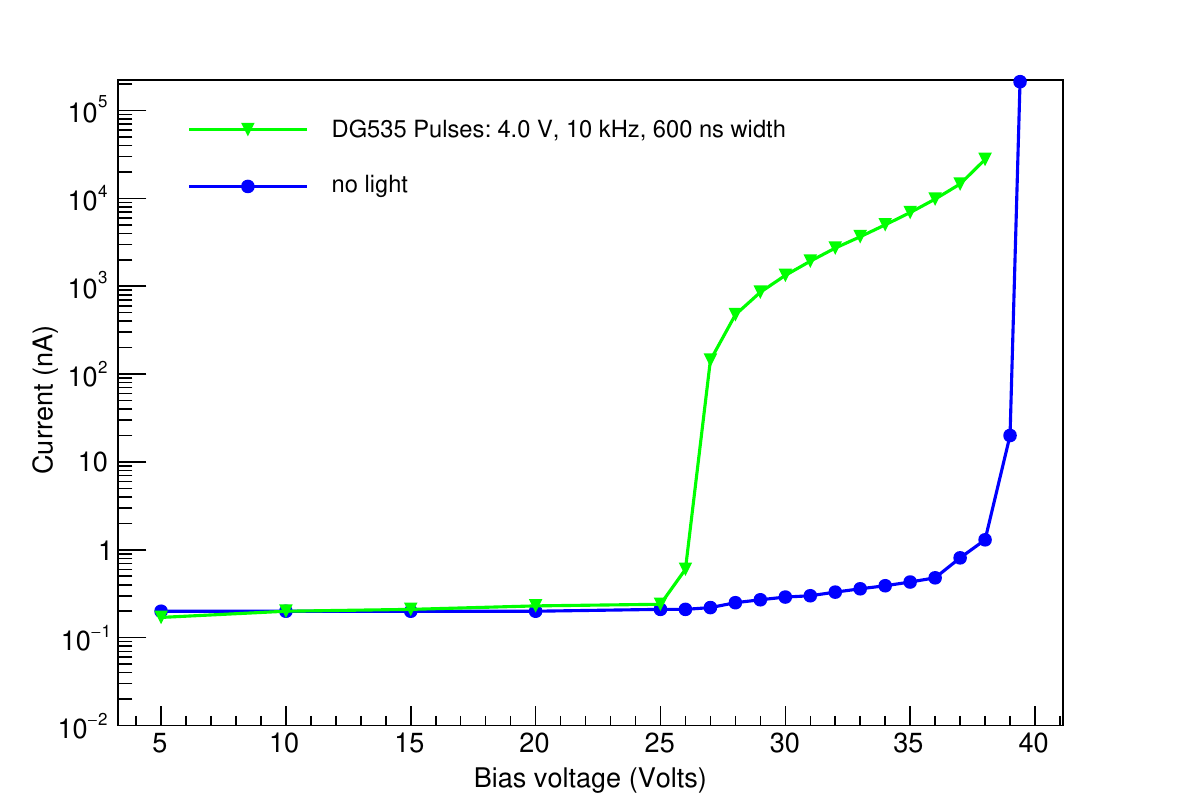}
	\caption{The typical I-V curve of an FBK NUV-HD-Cryo SiPMs measured at $\sim$ 10 K, with and without illumination. The green dots were measured with weak pulsed light from a 405 nm LED, which a DG535 drove. The blue ones were measured in the dark.}\label{schemeSiPMIVCurve10k} 
\end{figure}

We tested the FBK SiPMs' I-V curve at $\sim$ 4 K and RT in 2022~\cite{ALETHEIA-EPJP-2023}. The I-V curves we measured at RT are consistent with the measurements done in FBK laboratories at the same temperature. However, according to our communications with FBK, they were unaware that any group has ever tested their SiPMs' I-V curve near 4 K. So, we were not sure whether our tests were correct or not in 2022, even though the shape of the I-V curve seemed different from the one at RT: when the bias voltage is less than a threshold, the current is $\sim$ nA; as long as the applied voltage reaches the threshold, the current jumps to $\sim$ mA quickly, as the blue curve in Fig.~\ref{schemeSiPMIVCurve10k} shows. Although we repeated the I-V curve tests several times by different personnel, the same results were observed. The puzzle was solved in 2023, when the dark count rate (DCR) of the SiPMs was measured. Since the DCR was as low as 1 Hz at 10 K, the dark current (mainly composed of single electron events) was too small to build up a $\mu$A current, even if the bias was higher than the breakdown voltage~\footnote{Assuming the SiPMs' gain is 10$^6$ at 10 K with a bias higher than the breakdown voltage. Given the DCR is 1 Hz, the current would only be 0.16 pA = 1.6 $\cdot$ 10$^{-19}$ C $\times$ 1 Hz $\times$ 10$^6$, where 1.6 $\cdot$ 10$^{-19}$ C is the charge of an electron. At room temperature however, the DCR often is 10$^7$ Hz scale and the gain is 10$^6$, the current would be 1.6 $\mu$A = 1.6 $\cdot$ 10$^{-19}$ C $\times$ 10$^7$ Hz $\times$ 10$^6$.}. As long as the bias voltage reaches a threshold much greater than the breakdown voltage (25 V), 38 V ( = 13 V OV), for example, the correlated noise increases significantly and eventually generates $\sim$ mA current. The 1 Hz DCR is equivalent to $\sim$ 0.01 Hz/mm$^2$ for the $\sim$ 100 mm$^2$ SiPMs, which is 7 orders lower than the same photosensor at RT, $\sim$ 100 kHz/mm$^2$~\cite{Acerbi17}. 

Given the especially low DCR near LHe temperature, we tested the SiPMs' I-V curve in an alternative way: illuminating the SiPMs with weak pulsed light from an LED driven by a generator, DG535. During the I-V curve tests, the pulses used to drive the 405 nm LED were rectangular in shape, 4 V amplitude, 600 ns width, and 10 kHz frequency. Under such a testing setup, the I-V curve shows a conventional feature, as the green curve in Fig.~\ref{schemeSiPMIVCurve10k}. So far, we measured five randomly selected FBK SiPMs at 10 K with the same method, and all of the SiPMs recorded the same I-V curves. In addition, according to Fig.~\ref{schemeSiPMIVCurve10k}, the SiPMs' breakdown voltage at 10 K is around 25.0 V, which is consistent with other tests~\cite{Acerbi17, DarkSide20k17}, where 27.0 V and 32.0 V breakdown voltage were obtained on the similar SiPMs (NUV-HD-LF) at 40 K and RT, respectively. It is observed that the lower the temperature, the lower the breakdown voltage, possibly because a lower temperature would result in a higher electron ionization rate, as predicted in reference~\cite{Crowell68}.

\subsection{PDE measurements} \label{secPDETestAt10KsubsecMeasurements}

The PDE of certain types of FBK SiPMs were measured at RT in references~\cite{Zappala16, DarkSide20k17}. A $\ge$ 40\% PDE was obtained with the bias voltage of 5 V overvoltage (OV) or higher. However, according to our communications with FBK, their clients have yet to fully characterize the SiPMs' performance near LHe temperature. Being inspired by the PDE calibrations on SiPMs performed at LAr or LXe temperatures~\cite{Otte06, Finocchiaro09, Eckert10, Yang14, Piemonte16, Zappala16}, we characterized the PDE of the NUV-HD-Cryo FBK SiPMs near LHe temperature with 405 nm and 530 nm light.

\subsubsection{PDE tests at 10 K} \label{secPDETestSubSubsecAt10K}

We implemented the photocurrent method to measure the PDE, as introduced in reference~\cite{Zappala16}. The PDE is calculated as Eq.~(\ref{EqPDE}), 

\begin{equation}
PDE = \frac{\text{Measured\_current}}{\text{Incident\_current}} = \frac{ I_{SiPM-L} - I_{SiPM-D} } {G \times q_e \times ECF \times Ph_I}  \label{EqPDE},
\end{equation}
where $I_{SiPM-L}$ and $I_{SiPM-D}$ are the measured current of the SiPMs with the light on and off, respectively; $G$ is the gain of the SiPMs; $q_e$ is the charge of an electron; $ECF$ stands for Excess Charge Factor, which represents the fake efficiency prompt contributed from AP and CT~\cite{Piemonte12, MPPCTechnicalNote21}; $Ph_I$ is the incident photons.

Since a SiPMs' ECF is supposed to be the same with or without being illuminated, it can also be calculated when the SiPMs are in the dark,

\begin{equation}
ECF = \frac{ I_{SiPM} - I_{Leak} } {G \times q_e \times N_D} \label{EqECF},
\end{equation}
where $I_{SiPM}$ and $I_{Leak}$ represent the measured current for the bias voltage above or below the breakdown voltage in the dark, respectively; the definition of $G$ and $q_e$ are the same as Eq.~(\ref{EqPDE}); $N_\text{D}$ is the average number of photoelectrons per second in dark condition, which is equivalent to the average number of the Poisson statistical photoelectrons in a test~\cite{Otte06, Finocchiaro09, Eckert10, Yang14, Piemonte16, Zappala16}. As mentioned above in section~\ref{secPDETestAt10KsubsecIVCurve}, the SiPMs' DCR is as low as 1 Hz near LHe temperature in the dark, which leads to $\sim$ 0.1 pA current. While the measured current is $\sim$ 0.1 nA, essentially the detector system's dark current. So, we illuminated the SiPMs with a weak pulsed light for the ECF tests to have a $\mu$A scale current. Effectively, our testing condition of  ``weak light at LHe temperature'' is equivalent to the conventional one of ``dark at RT''.

Substituting Eq.~(\ref{EqECF}) into Eq.~(\ref{EqPDE}), one gets the PDE expression as Eq.~(\ref{EqPDESubstituted}).

\begin{equation}
PDE =  \frac{ N_D \times (I_{SiPM-L} - I_{SiPM-D}) } { (I_{SiPM} - I_{Leak} ) \times Ph_I} \label{EqPDESubstituted},
\end{equation}

\begin{figure}[!t]	 
	\centering
    \includegraphics[width=0.8\textwidth]{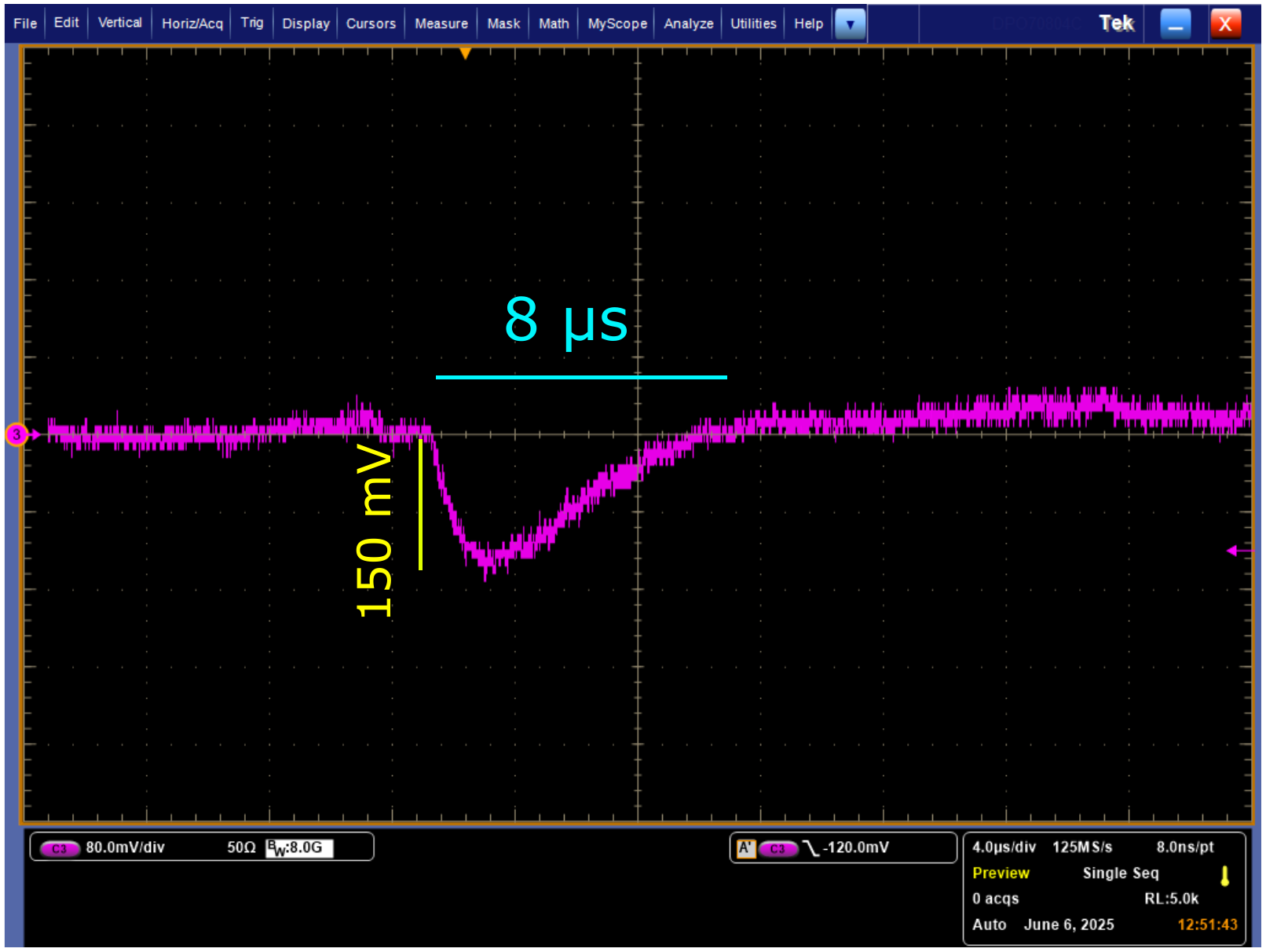}
	\caption{A typical single Phe signal of the FBK SiPMs measured around 10 K. According to our measurements with the V1 preamplifier, the SiPM signal has a typical rise time on the nanosecond scale at 10 K. The 2 $\mu$s rise time shown in the figure is not intrinsic to the SiPMs but is instead limited by the 170 kHz bandwidth of the V2 preamplifier used for that measurement.}\label{typicalRawSignal10K} 
\end{figure}

The SiPMs were illuminated with 405 nm or 530 nm light from an LED driven by a DG535 pulse generator. For PDE tests, the pulses from the DG535 are 4 V amplitude, 240 ns width, and 10 kHz frequency. Whenever the DG535 generated a pulse, it also triggered an 8 GHz bandwidth, 25 GSa/s sample rate oscilloscope to record the waveform showing on the screen. A typical single photoelectron (Phe) signal of the FBK SiPMs at 10 K is shown in Fig.~\ref{typicalRawSignal10K}. Given that (a) the effective time window of the pulses is 2.4e-3 s (= 240 ns $\times$ 10 kHz) per second, and (b) the DCR is about 1 Hz near LHe temperature, the dark current events only have a chance of 0.24\% (= 2.4e-3 s / 1 s) to show up in data. Therefore, the DCR will not smear the PDE results.

\begin{figure}[!t]	 
	\centering
    \includegraphics[width=0.8\textwidth]{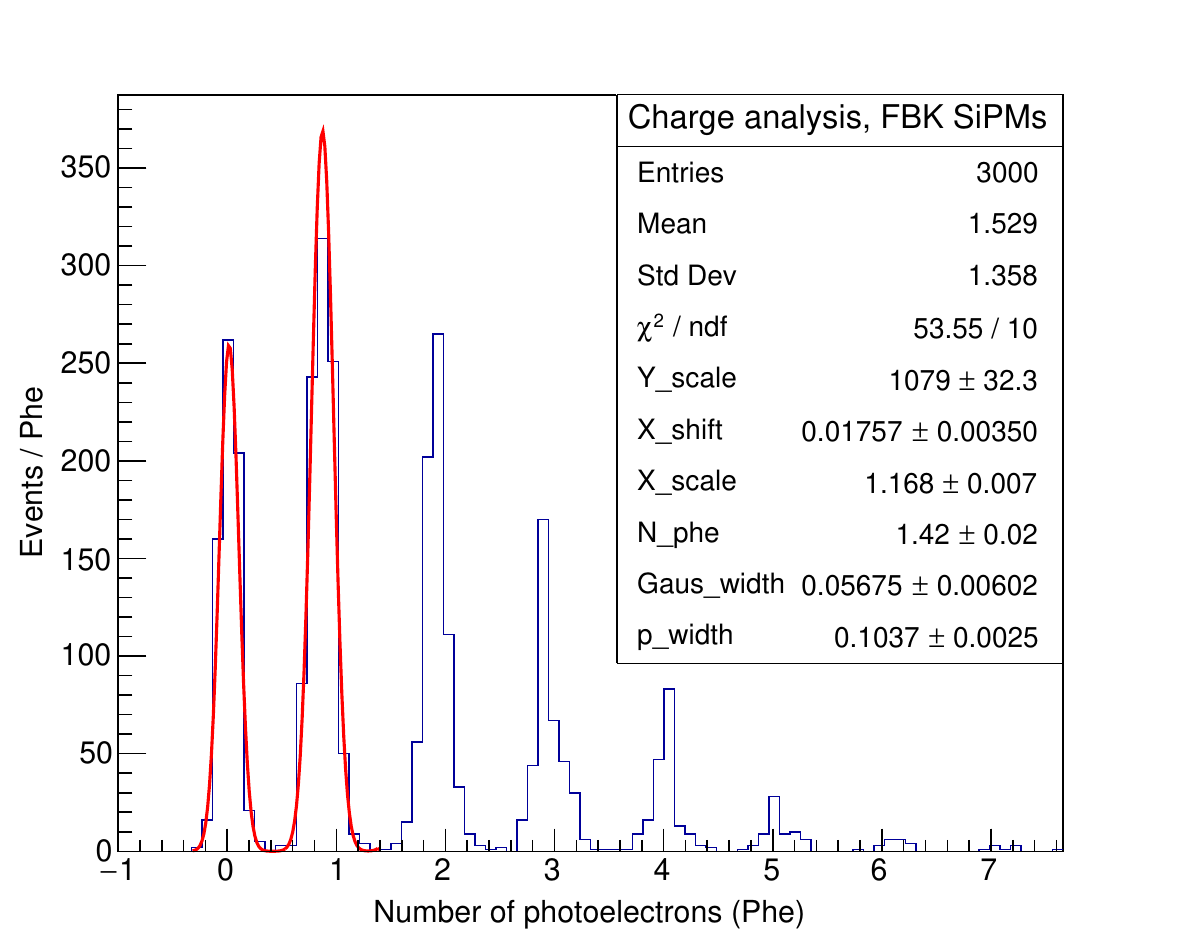}
	\caption{A typical charge distribution we measured around 10 K for an FBK NUV-HD-Cryo SiPMs being biased with OV = 6 V (31 V). The fit function is Gaussian convoluted with Poisson~\cite{Gastof14}. The fit was intentionally applied on the 0 and 1 Phe peaks only since the 2+ Phe peaks contained CT smearing events. For more info on the figure, please refer to the main text.}\label{SiPM10K35VQFit} 
\end{figure}

As mentioned above, we must figure out the average number of photoelectrons to calculate PDE. We get it (mainly) through a fit on a charge histogram as shown in Fig.~\ref{SiPM10K35VQFit}. The histogram was obtained by integrating thousands of analog signals in a $\sim$10 $\mu$s window.  Fig.~\ref{typicalRawSignal10K} shows such a typical signal. During data-taking, the DG535 was tuned to generate weak enough light to produce $\sim$ 1.5 Phe in the SiPMs on average. The oscilloscope was triggered by the DG535.  The signals were firstly recorded in the oscilloscope, then analyzed offline with ROOT~\cite{rootWebsite}-based scripts. A recorded raw signal often contained roughly 2000 data points, which were integrated directly (without baseline subtraction), then divided by a 50 ohm (the input resistance of the oscilloscope) to convert into charge. Next, we multiplied the charge with the system's gain to get the  number of Phe, as shown in Eq.~(\ref{tentativePheCal}),

\begin{equation}
\text{Number of Phe} =  \frac{raw ~signal ~integral}  {50 ~\Omega \times gain \times 1.6 \cdot 10^{-19}} \label{tentativePheCal},
\end{equation}
where the $raw ~ signal ~ integral$ is the integral of raw signals, 50 $\Omega$ is the input resistance of the oscilloscope, the $gain$ is the gain of the SiPMs (often is the scale of 1.0 $\cdot$ 10$^{6}$) and the preamplifier combined, 1.6 $\cdot$ 10$^{-19}$ is the charge of an electron. 
With the same setup and analysis, under the OV between 5 and 11 V, all of the charge histograms have shown explicit separated Phe peaks. Fig.~\ref{SiPM10K35VQFit} shows a typical charge distribution with OV = 6 V and a fit. The fit function is Gaussian convoluted Poisson, as shown in Eq.~\eqref{fitFucntion}. 

\begin{align}
pedestal &= e^{-\mu}  [Gauss((x - x_{shift}) \cdot x_{scale}, 0, p\_{width} )] \label{pedestal} \\
nPhePeaks &= \sum_{n = 1}^N \frac{e^{-\mu}} {n!} \mu^n  [Gauss (x - x_{shift}) \cdot x_{scale}, n, \sqrt{n \cdot  (Gaus\_{width})^2 + (p\_{width})^2} ~] \label{nPhePeaks} \\
fitFunction &= Y_{scale} ~ (pedestal + nPhePeaks ) \label{fitFucntion}
\end{align}

The meaning of the parameters in the above equations are the following. ``$pedestal$'', ``$nPhePeaks$'', and ``$fitFunction$'' are the fit functions correspond to the pedestal (or 0 Phe), 1 to N Phe peaks, and the whole histogram, respectively. ``$\mu$'' is the average number of the Poisson function. ``$Y\_scale$'' means the scale factor of the fit function on the Y-axis. ``$X\_shift$'' indicates the shift on the X-axis, which is the analog signals' baseline shift on the oscilloscope. ``$X\_scale$'' corresponds to the scale factor on the X-axis. ``$Gaus\_width$'' and ``$p\_width$'' are the 1-$\sigma$ of the one Phe peak and the pedestal distribution, respectively.

\begin{figure}[!t]	 
	\centering
    \includegraphics[width=0.8\textwidth]{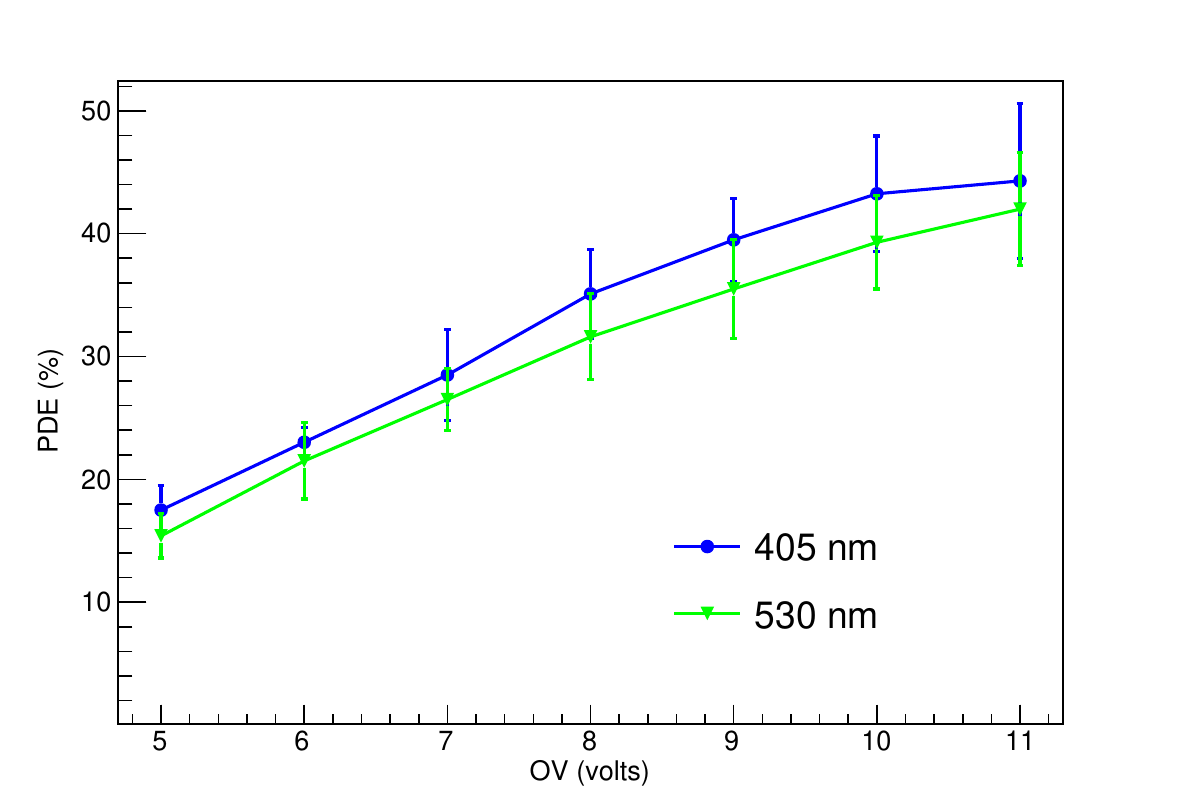}
	\caption{The measured PDE for an FBK NUV-HD-Cryo SiPMs working at 10 K under variant bias voltages illuminated with 405 nm and 530 nm light. The PDE equals or exceeds 40\% at $\ge$ 10 V OV.}  \label{allPDESiPM10K} 
\end{figure}

In Fig.~\ref{SiPM10K35VQFit}, we limited our fit function on the less than 2 Phe part of the histogram to avoid smearing the fit by AP and CT events~\cite{Acerbi17, AcerbiGundacker19}. The average number of photoelectrons, 1.42 $\pm$ 0.02 Phe, is the N$_D$ to be substituted in Eq.~(\ref{EqPDESubstituted}) for PDE calculation. As will be discussed further in~\ref{secTestAt10KSubSubsecCT}, the 1.42 $\pm$ 0.02 is consistent with the ratio of the events in the ``0 Phe'' and ``1 Phe'' peaks assuming the events follow a Poisson distribution, therefore, justifying the reliability of the fit. The events in the ``0 Phe'' and ``1 Phe'' peaks are 672 and 963, respectively; and 672 / 963 $\approxeq$ Poisson (0, 1.42) / Poisson (1, 1.42), where Poisson (0, 1.42) and Poisson (1, 1.42) are the probabilities of 0 and 1 under a Poisson distribution with 1.42 on average, respectively. Other parameters we measured or calculated for the PDE test are shown in table~\ref{tabAllParameters}. Submitting these parameters into Eq.~(\ref{EqPDESubstituted}) gives the PDE of (23.02 $\pm$ 1.18)\%. Following the same procedure, the PDE values measured with other bias voltages and 530 nm light can also be obtained, as shown in Fig.~\ref{allPDESiPM10K}. From the plot, we can see that for a bias voltage of 10 V OV or more,  the PDE is equal to or greater than 40\%.

\newcommand{\otoprule}{\midrule[\heavyrulewidth]} 

\begin{table}[ht]
   \centering
   \caption{All of parameters were used for a PDE calculation, in which the SiPMs were illuminated by 405 nm light, biased with 31 V (OV = 6 V) bias voltage, and operated at 10 K.} \label{tabAllParameters}   
      \begin{tabular}{c c c c c c c}
      \toprule%
         \multicolumn{1}{c}{\bfseries{ N$_D$} }  &
         \multicolumn{1}{c}{\bfseries{ I$_{SiPM-L}$} }  &
         \multicolumn{1}{c}{\bfseries{I$_{SiPM-D}$ } }  &
         \multicolumn{1}{c}{\bfseries{I$_{SiPM}$ } }  &
         \multicolumn{1}{c}{\bfseries{I$_{Leak}$ } }  &
         \multicolumn{1}{c}{\bfseries{Ph$_I$} }     &%
         \multicolumn{1}{c}{\bfseries{PDE} }     \\%
	(Phe)			&($\mu$A)	&(nA)			&(nA)	&(nA)	&(photon) 	&(\%)\\
	  \otoprule%
      1.42$\pm$0.02	&4.77$\pm$0.09	&0.12$\pm$0.07	&2.64$\pm$0.02	&0.02$\pm$0.01	&(6.83$\pm$0.31)E6 		&23.02$\pm$ 1.18 \\
      \bottomrule
    \end{tabular}
\end{table}

\subsubsection{Gain tests at 10 K} \label{secGainTestSubSubsecAt10K}

\begin{figure}[!t]	 
	\centering
    \includegraphics[width=0.8\textwidth]{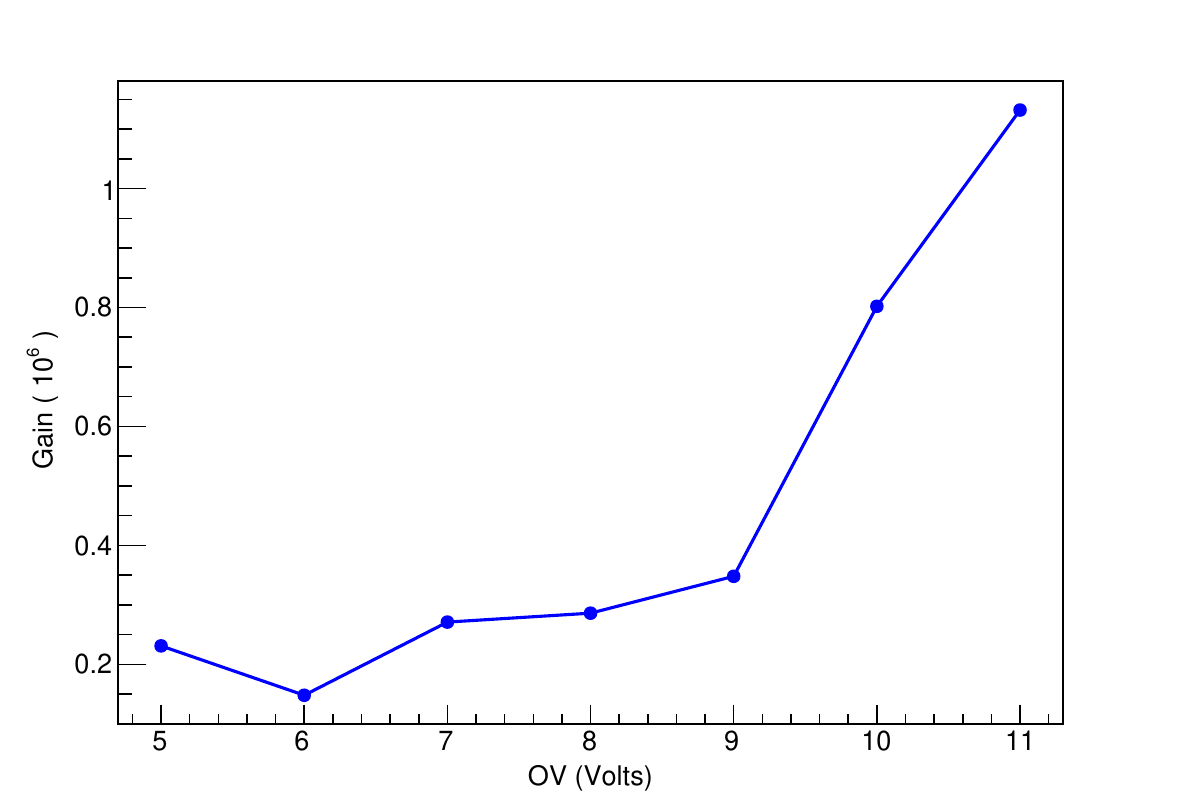}
	\caption{The measured gain of an FBK SiPMs at 10 K, being illuminated by 405 nm light. For more info, please refer to the main text.}\label{FigGain10K} 
\end{figure}

We tested the SiPMs' gain at 10 K with 405 nm light, as shown in Fig.~\ref{FigGain10K}. With the same bias voltage of OV = 5 V, the gain of the NUV-HD-Cryo SiPMs at 10 K was 0.23 $\times$ 10$^6$; while the gain of the NUV-HD-LF type of SiPMs at 40 K was $\sim$ 0.32 $\times$ 10$^6$~\cite{Acerbi17, DarkSide20k17}. Our results are consistent with previous measurements, which observed the feature of gain decreasing with temperature. However, we identified a difference in the relationship between gain and overvoltage. Tests on the same type of SiPMs at liquid nitrogen temperature reported a linear relationship between 2 - 10 V OV~\cite{Acerbi17, Capasso20}, a feature not observed in our tests (Fig.~\ref{FigGain10K}). Given that a non-linear relationship was also observed in an independent test on a different type of SiPMs at 100 mK~\cite{SiPMsTests100mk}, we suspect the linear relationship may not hold for temperatures below 10 K, although a convincing model is still lacking. We will continue to investigate this issue in future tests.

\subsection{DCR, AP, and CT tests at 10 K}\label{secDCRAPCTTestAt10K} 
The dark count rate (DCR), after-pulse (AP), and cross-talk (CT) are critical parameters for SiPMs. Although these parameters were measured for two types of FBK SiPMs from RT down to 40 K with a different DAQ system, as shown in reference~\cite{Zappala16, Acerbi17}, we must characterize them near LHe temperature again for the following reasons. (a) The SiPMs under our tests are not the same model as the references according to our discussion with FBK engineers, though it is similar to the NUV-HD-LF one. (b) The dimension of our SiPMs is 11.7 $\times$ 7.9 mm$^2$, which is a factor of six greater than the ones tested in reference~\cite{Zappala16, Acerbi17}, 4.0 $\times$ 4.0 mm$^2$; the cell pitch of our SiPMs is 30 $\times$ 30 $\mu$m$^2$, which is also bigger than the 25 $\times$ 25 $\mu$m$^2$ one in the references. (c) For FBK SiPMs, the lowest temperature the DCR, AP, and CT have been tested was 40 K, while the working temperature of the SiPMs equipped on LHe detectors is supposed to be near LHe temperature; so, we need to test these parameters near $\sim$ 4.5 K.

\subsubsection{DCR tests at 10 K}\label{secTestAt10KSubSubsecDCR} 
The setup for the DCR tests is nearly identical to the PDE tests, as shown in Fig.~\ref{integratingSphereOnCoolingPlateRealPic}. However, the electronics are slightly different: a 5$\times$ amplifier was connected to the preamplifier, and a counter read the signals coming out of the amplifier. Another difference was that the LED was off during the DCR tests. The counter was in a self-trigger model: once the signal's amplitude crosses the counter's threshold, the signal will be registered, and the count number will increase by one accordingly. 

\begin{figure}[!t]	 
	\centering
    \includegraphics[width=0.8\textwidth]{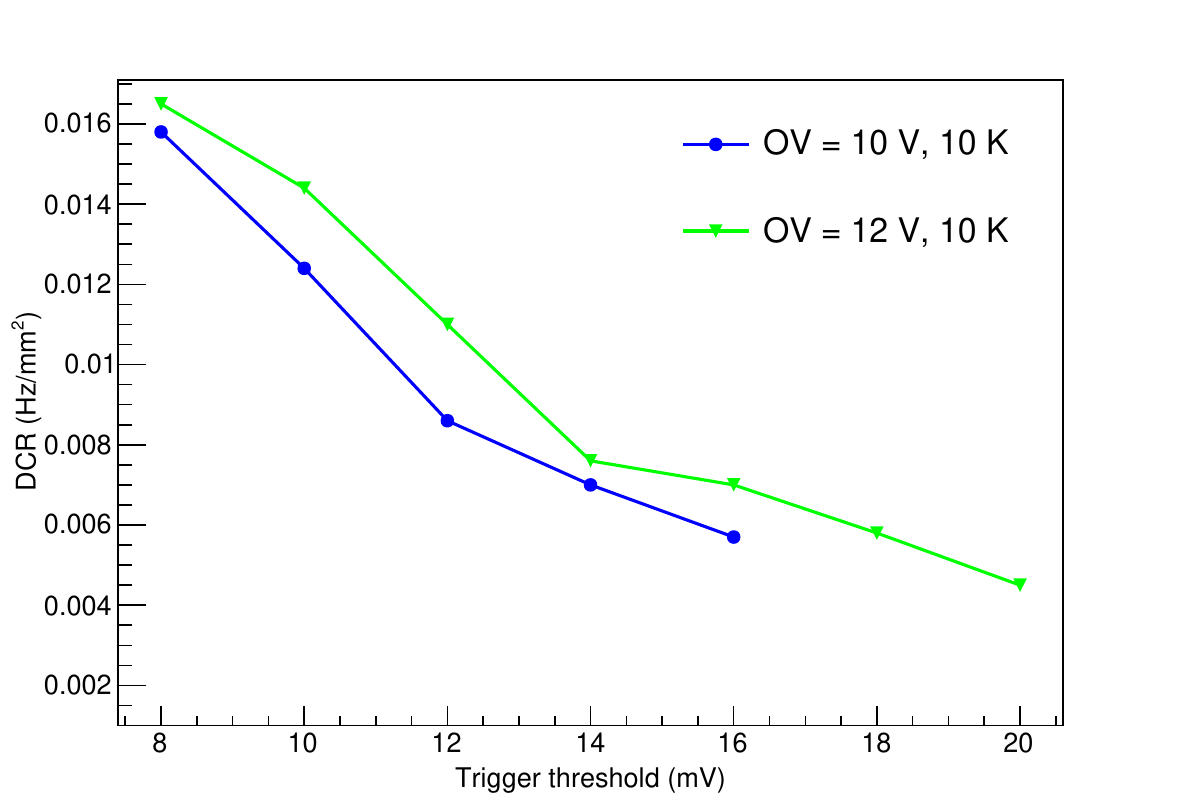}
	\caption{The DCR of an FBK NUV-HD-Cryo type SiPMs at 10 K with the bias voltage of OV 10 V and 12 V. The SiPMs' dimension is 11.7 $\times$ 7.9 mm$^2$ with the cell pitch of 30 $\times$ 30 $\mu$m$^2$. The DCR curve lacks the characteristic ``step'' shape seen in other tests like Ref.~\cite{Eckert10}, because the measurement was performed using a V1 preamplifier, which has limited charge resolution.}\label{SiPMDCR10k} 
\end{figure}

For the NUV-HD-Cryo type SiPMs under test, Fig.~\ref{SiPMDCR10k} shows the measured DCR is $\sim$ 0.01 Hz/mm$^2$ for OV 10 V and 12 V at 10 K. Reference~\cite{Acerbi17} measured the DCR for two types of FBK SiPMs: NUV-HD-SF and NUV-HD-LF. At RT, the typical DCR with 4 to 6 V OV is $\sim$ 100 kHz, while at 40 K, the DCR is 0.1 Hz/mm$^2$ and 0.001 Hz/mm$^2$ for the NUV-HD-SF and NUV-HD-LF SiPMs, respectively. 
Given that (a) the dimension of our SiPMs is $\sim$ 6 times that of the ones tested in the reference, and (b) the OV in our tests (10 to 12 V) is 6 volts higher than the paper (4 to 6 V), we consider our DCR results are consistent with the NUV-HD-LF ones measured in the paper at 40 K. We stick to a 10 V OV instead of a lower bias because it is the voltage the SiPMs can reach a required PDE criterion for LHe TPCs, 40\%.

\subsubsection{AP tests at 10 K}\label{secTestAt10KSubSubsecAP}

\begin{figure}[!t]	 
	\centering
    \includegraphics[width=0.8\textwidth]{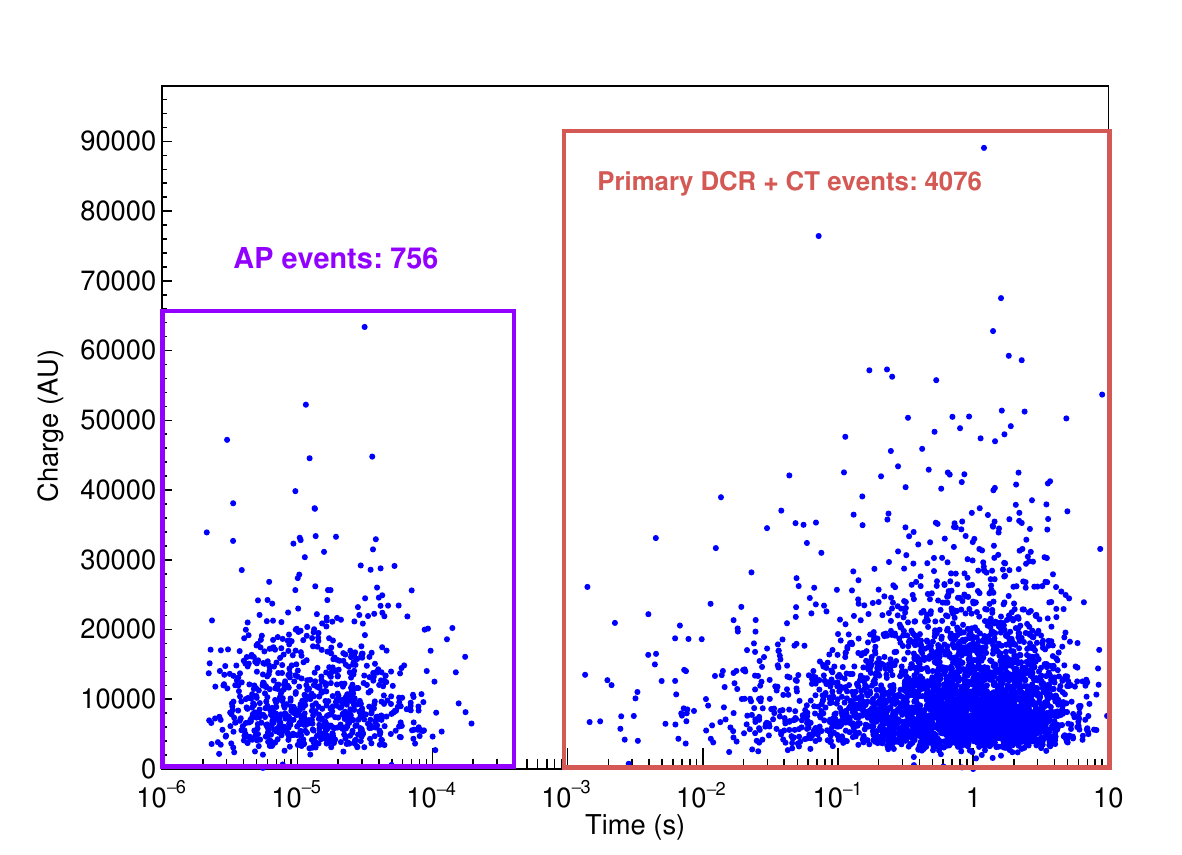}
	\caption{The measured AP, primary DCR, and CT events for an FBK SiPMs biased with 10 V OV (i.e., 35 V) at 10 K. }\label{SiPMAPDCR10k} 
\end{figure}

The setup for the AP tests is almost the same as the DCR measurements (the LED was off either); the only difference is that the counter was replaced with a digitizer, the DAQBOX-4-125 model, from TOFTEK company in China. The digitizer has a 125 MSPS sampling rate, a 12-bit resolution, and an input range of -1 V to +1 V. It operates in an auto-trigger mode: whenever a signal's amplitude crosses a programmable threshold, the timing crossing the threshold will be registered as the pulse's time, T$_0$. A user-definable time interval will be used to define a time window. Assuming the interval is set to be 1 second, the time window would be (T$_0$, T$_0$ + 1)s. And all of the pulses within the window will be registered. For each recorded pulse, the time is the difference between its threshold-crossing timing and the T$_0$; the charge is the integration of the analog signal (by the digitizer). Fig.~\ref{SiPMAPDCR10k} shows the time and charge info of the signals recorded in 10 seconds. During data-taking, the threshold was set to be 40 mV~\footnote{The 40 mV was measured with a 5$\times$ amplifier connected upstream. So, it is equivalent to 8 mV on the preamplifier's output.}, and the time window of each pulse was 2 $\mu$s. The 2 $\mu$s time window is wide enough for the $\sim$ 1.5 $\mu$s width signals \footnote{The preamplifier we implemented in the AP tests is different from the one    being utilized for PDE and CT tests. Among other difference, the former has a 1.5 $\mu$s integration time, while the latter is $\sim 8 \mu$s, as shown in Fig.~\ref{typicalRawSignal10K}.} and narrow enough to exclude unnecessary noises.
As shown in Fig.~\ref{SiPMAPDCR10k}, there are 768 AP events and 4076 primary DCR and CT events. The AP ratio is calculated to be 768 / (768 + 4076) = 15.9\%. 

\begin{figure}[!t]	 
	\centering
    \includegraphics[width=0.8\textwidth]{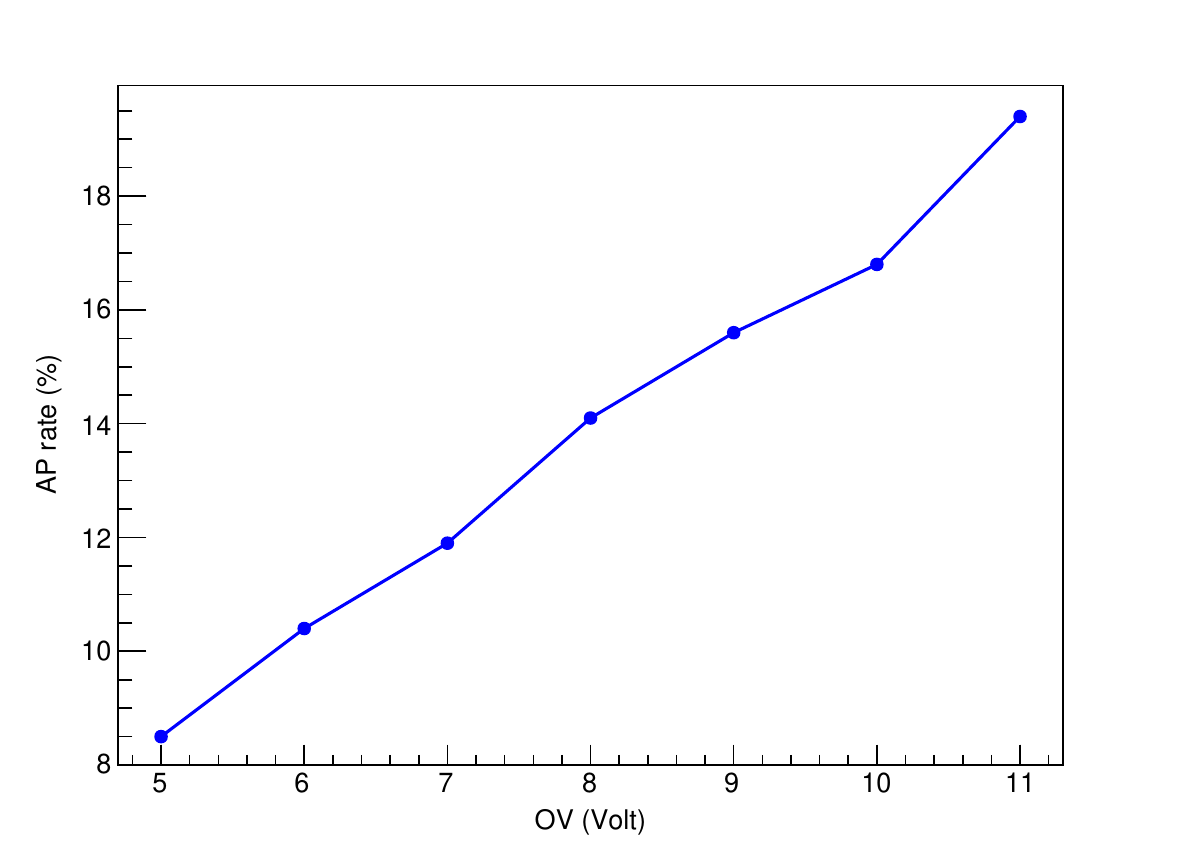}
	\caption{The AP rate of the FBK SiPMs working at 10 K when biased with 10 V OV (35 V).}\label{SiPMAPVSBias10k} 
\end{figure}

At 10 K, we also tested the AP under other bias voltages, as Fig.~\ref{SiPMAPVSBias10k} shows. The measured AP rate at 10 K is (surprisingly) consistent with the results at 40 K~\cite{Acerbi17} for the NUV-HD-LF type SiPMs in the sense that (a) the relationship between the bias voltage and the AP rate is the same: the higher the bias, the greater the AP; and (b) with the same bias voltage of OV 5 V or 6 V, the measured rate are both around $\sim$ 10\%.

\subsubsection{CT tests at 10 K}\label{secTestAt10KSubSubsecCT}

The data for the CT analysis was the same as for the charge analysis when the SiPMs was illuminated with weak light, as shown in Fig.~\ref{SiPM10K35VQFit}. The CT analysis can be split into three steps: (a) Figuring out the average number of photoelectrons, $\mu$, based  on the number of events in ``0 Phe'' and ``1 Phe'', which correspond to the first and second peaks in Fig.~\ref{SiPM10K35VQFit}, respectively; (b) Extrapolating the expected number of events in the histogram, N$_{\text{expect}}$, under the reasonable assumption of no CT events included; (c) calculating the CT rate with the equation of (N$_{\text{all}}$ - N$_{\text{expect}}$) / N$_{\text{all}}$, where N$_{\text{all}}$ is the charge histogram's total events, which is 3000 in Fig.~\ref{SiPM10K35VQFit}. In the figure, the events in the `0 Phe'' and ``1 Phe'' peaks are 672 and 963, respectively. Assuming these events follow a Poisson distribution, the $\mu$ of the histogram would be 1.43 since 672 / 963 $\approxeq$ Poisson (0, 1.43) / Poisson (1, 1.43),  while the fit result is 1.42 $\pm$ 0.02, perfectly lines up with the calculation based on the events in the two peaks. In addition, since all the charge histograms we measured with OV between 5 and 11 V have explicitly separated Phe peaks, which means there are no smearing events across the peaks. As a result, for the CT analysis, we rely on the events in the peaks (instead of the fit results).

According to the events in the ``0 Phe'' (672) and ``1 Phe'' (963) peaks, the average number Phe in Fig.~\ref{SiPM10K35VQFit} is 1.43 since 672 / 963 $\approxeq$ Poisson (0, 1.43) / Poisson (1, 1.43), so the expected events, N$_{\text{expect}}$, can be obtained as the following Eq.(\ref{ExpectedEvents}) provided no smearing events in the 2+ peaks.

\begin{equation}
{\text{N}}_{\text{expect}}  = 963 \times \frac {Poisson (2, 1.43)}{Poisson (1, 1.43)} +  ... + 963 \times \frac {Poisson (7, 1.43)}{Poisson (1, 1.43)} = 1177  \label{ExpectedEvents},
\end{equation}

The observed events in all the 2 + peaks in Fig.~\ref{SiPM10K35VQFit} are 1362. So the discrepancy, 1362 - 1177 = 185,  is the number of CT events. And the preliminary CT ratio is 185 / 3000 = 0.0615 = 6.15\%. Since (a) a CT event happens only when one photon emits, and (b) the mean Phe are not exactly the same in each dataset with different OV values because the LED's light intensity decreases along with the running time around 10 K according to our tests, so we normalized the CT ratio corresponding to 1 Phe by dividing the preliminary one with $\sqrt{1.43}$ to get the final value, 5.15\%. In uncertainty estimation, we simplified it by only considering the errors in the final step of calculation, i.e, the preliminary ratio. The errors for 185 and 3000 are $\sqrt{185}$ (= 13.59) and $\sqrt{3000}$ (= 54.77), respectively. We then propagate the errors according to the calculation ( (185 / 3000) / $\sqrt{1.43}$ ) to get the final uncertainty, 6.33\%. And the final CT ratio is 5.15\% $\pm$ 6.33\%.

\begin{figure}[!t]	 
	\centering
    \includegraphics[width=0.8\textwidth]{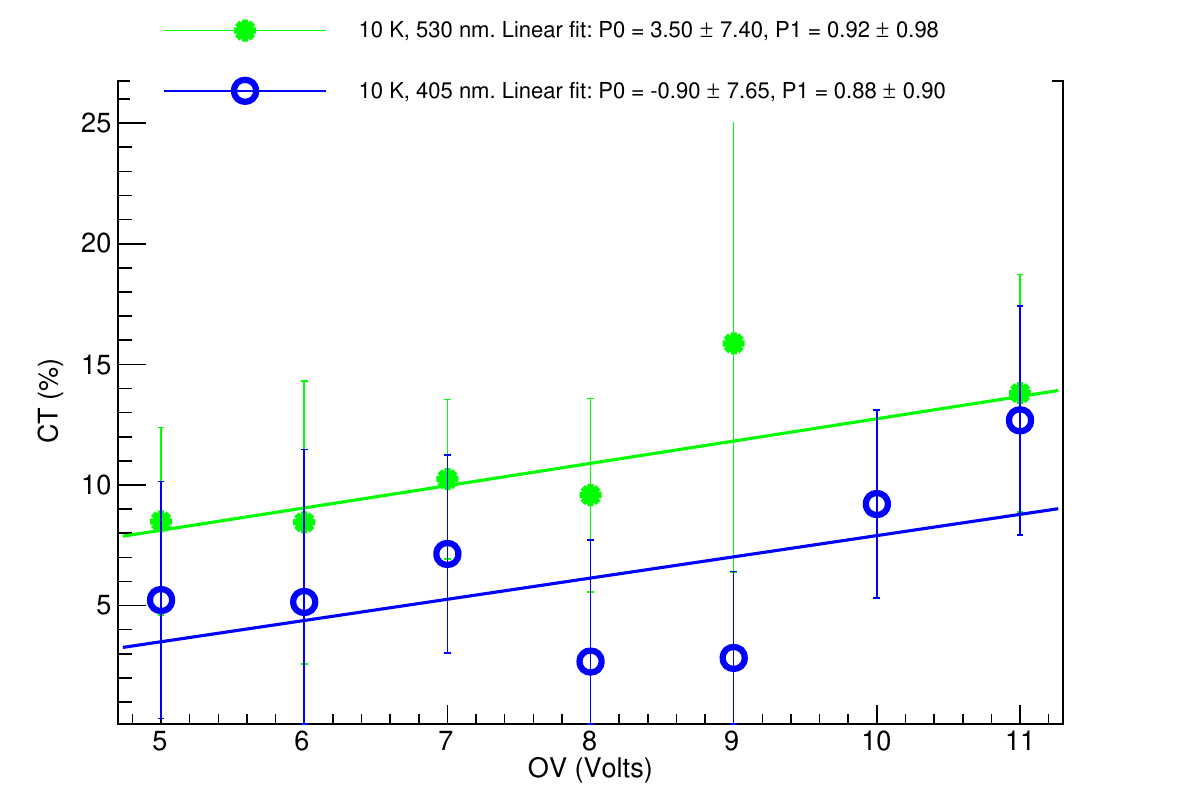}
	\caption{The CT rate for the FBK SiPMs operating at 10 K, illuminated with 405 and 530 nm light and biased with an OV from 5 V to 11 V. The linear fit results on the CT rate data are shown on the top of the plot. P0 is the constant, and P1 is the slope. For more info on the analysis, please refer to the main text.}\label{SiPMsCT7K10KOV5To11V} 
\end{figure}

We measured the SiPMs' CT being biassed with 5 to 11 V OV and calculated the uncertainties, as summarized in Fig.~\ref{SiPMsCT7K10KOV5To11V}~\footnote{Be aware please, the CT analysis we did here is different from others, eg, Ref.~\cite{Eckert10}.}. In the plot, all of the CT rate values and their uncertainties were normalized to have 1.0 Phe. Further, a linear fit was applied to fit the CT rate data. P0 is the constant, and P1 is the slope. We saw the expected trend of the higher the bias, the greater the CT rate.

\section{Understanding the SiPMs' temperature} \label{secUnserstandSiPMsTemp}

\begin{figure}[!t]	 
	\centering
    \includegraphics[width=0.8\textwidth]{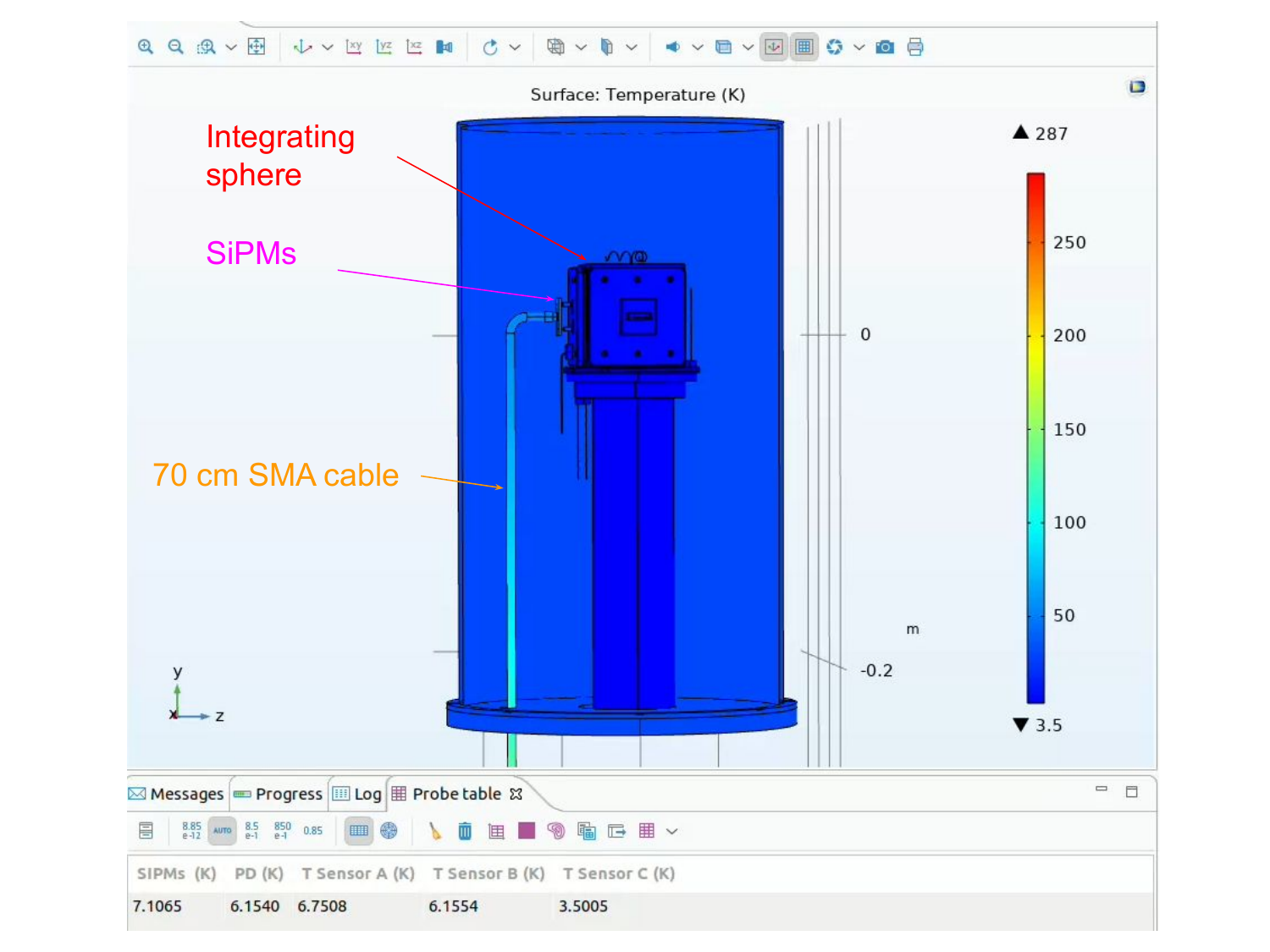}
	\caption{A COMSOL simulation shows that the SiPMs and the temperature sensor are 7.1065 K and 6.7508 K, respectively. ``PD'' refers to the photodiode. ``T Sensor B'' is the temperature sensor near the PD, and ``T Sensor B'' is the temperature sensor on the cooling plate.}\label{SiPMsCOMSOL7K} 
\end{figure}

In parallel to testing the SiPMs' performance at 10 K, we launched a COMSOL~\cite{comsolWebsite} simulation to crosscheck whether the 10 K temperature measured by the sensor makes sense. In the simulation, all the materials  and specifications are precisely the same as in the experimental setup; other input parameters are either directly measured or extrapolated from the measured data in situ. 
The simulation results are shown in Fig.~\ref{SiPMsCOMSOL7K}. The temperature sensor near the SiPMs was 6.75 K, as the ``T Sensor A (K)'' indicates; while the SiPMs' temperature was 7.1 K. That being said, there was a 0.35 K discrepancy between the temperature sensor and the SiPMs even though they have been mounted on the same copper plate and only have $\sim$ 5 cm distance. According to our experience in simulating other setups, the temperature discrepancy should be $\le$ 0.1 K, given the two parts' configuration. The ``abnormal'' discrepancy occurred because the SiPMs are connected to a 70 cm SMA cable, through which a preamplifier connected on another end at RT. As a result, the SiPMs has an extra ``warm'' source from the cable, in addition to the cryogenic source from the integrating sphere. Even by eye, we can see in Fig.~\ref{SiPMsCOMSOL7K} that the temperature on the cable's cold end (near the SiPMs) is higher than the integrating sphere. In the figure, ``PD'' refers to the photodiode. ``T Sensor B'' is the temperature sensor near the PD, and ``T Sensor C'' is the temperature sensor on the cooling plate, which is the cryogenic source of the detector system. 

\begin{figure}	
	\centering
	\begin{subfigure}[t]{2.5in}
		\centering
		\includegraphics[width=1.1\textwidth]{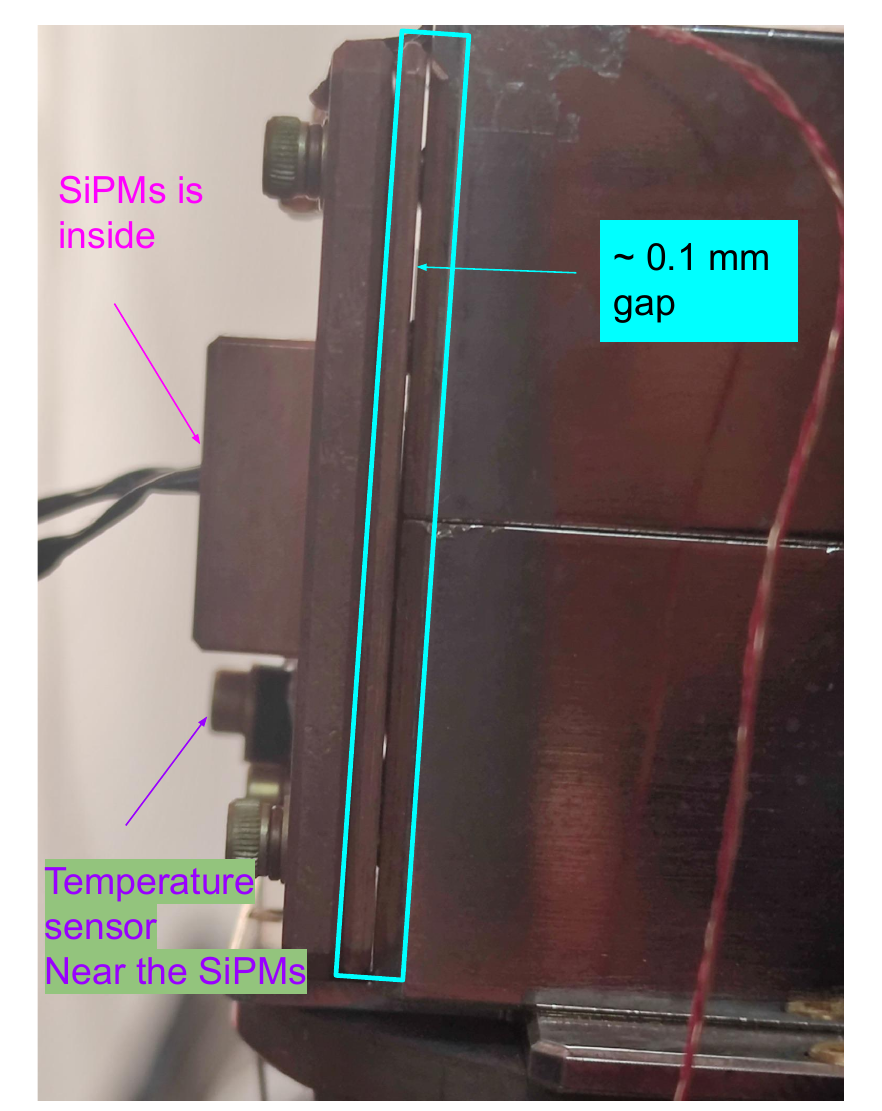}
		\caption{Fig.~\ref{figIntegratingSphereGap.a}. A $\sim$ 0.1 mm gap existed between the covering plate and the integrating sphere's wall. The gap made the temperature of the SiPMs and its temperature sensor nearby $\sim$ 3 K higher than it should be.} \label{figIntegratingSphereGap.a}	
	\end{subfigure}
	\quad
	\begin{subfigure}[t]{2.5in}
		\centering
		\includegraphics[width=1.1\textwidth]{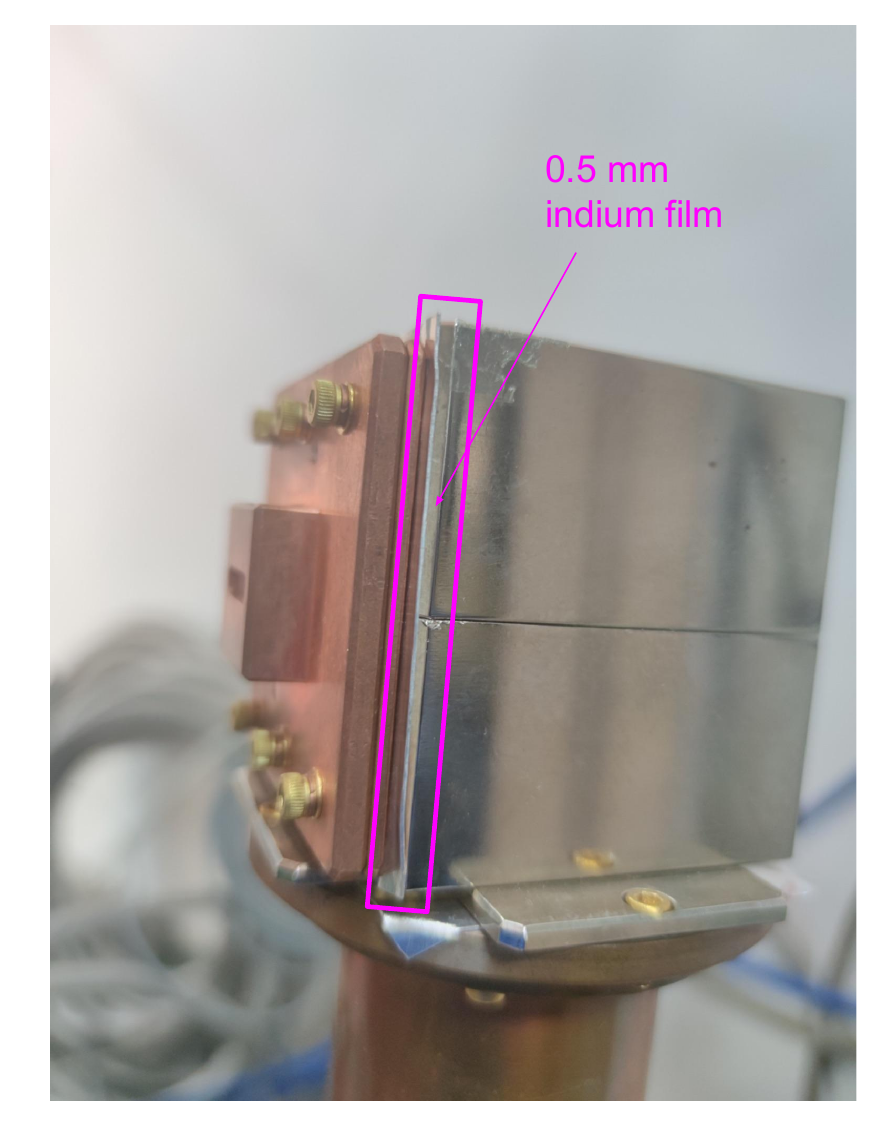}
		\caption{Fig.~\ref{figIntegratingSphereGap.b}. The gap was filled with a 0.5 mm-thick indium film. This update led to the measured temperature of the sensor dropping by approximately 3.0 K down to 7.0 K, which aligned well with the simulation in Fig.~\ref{SiPMsCOMSOL7K}, 6.75 K, confirming the accuracy of our simulation.} \label{figIntegratingSphereGap.b}
	\end{subfigure}
	\caption{An unexpected $\sim$ 0.1 mm gap led to the temperature of the SiPMs 3 K higher than expected. And a 0.5 mm-thick indium film was inserted to fix the problem.}\label{figIntegratingSphereGap}
\end{figure}

The simulation suggested that the temperature sensor should be 6.75 K. In comparison, the measured one was 10 K, and the $\sim$ 3 K ( 10 K - 6.75 K) discrepancy is greater than the uncertainty of the temperature sensor, $\pm 1$ K. So, we suspected there should exist other unknown reason(s) being responsible to the temperature discrepancy. We carefully checked the experimental setup and surprisingly spotted a $\sim$ 0.1 mm gap between the copper plate (in which the SiPMs mounted) and the integrating sphere wall, as Fig.~\ref{figIntegratingSphereGap.a} shows. We found similar gaps on the other sides of the cubic body. The gaps were due to stress relief during the cooling-down and warming-up cycles. We inserted a 0.5 mm-thick indium film between the plate and the sphere wall, as shown in Fig.~\ref{figIntegratingSphereGap.b}. After the upgrade, the temperature sensor near the SiPMs decreased to 7.0 K, 3 K lower than before and consistent with the simulation.

\captionsetup[subfigure]{labelformat=empty}

\section{Characterizing the SiPMs at 7 K} \label{secCharacterization7K}

\begin{figure}[!t]	 
	\centering
    \includegraphics[width=0.8\textwidth]{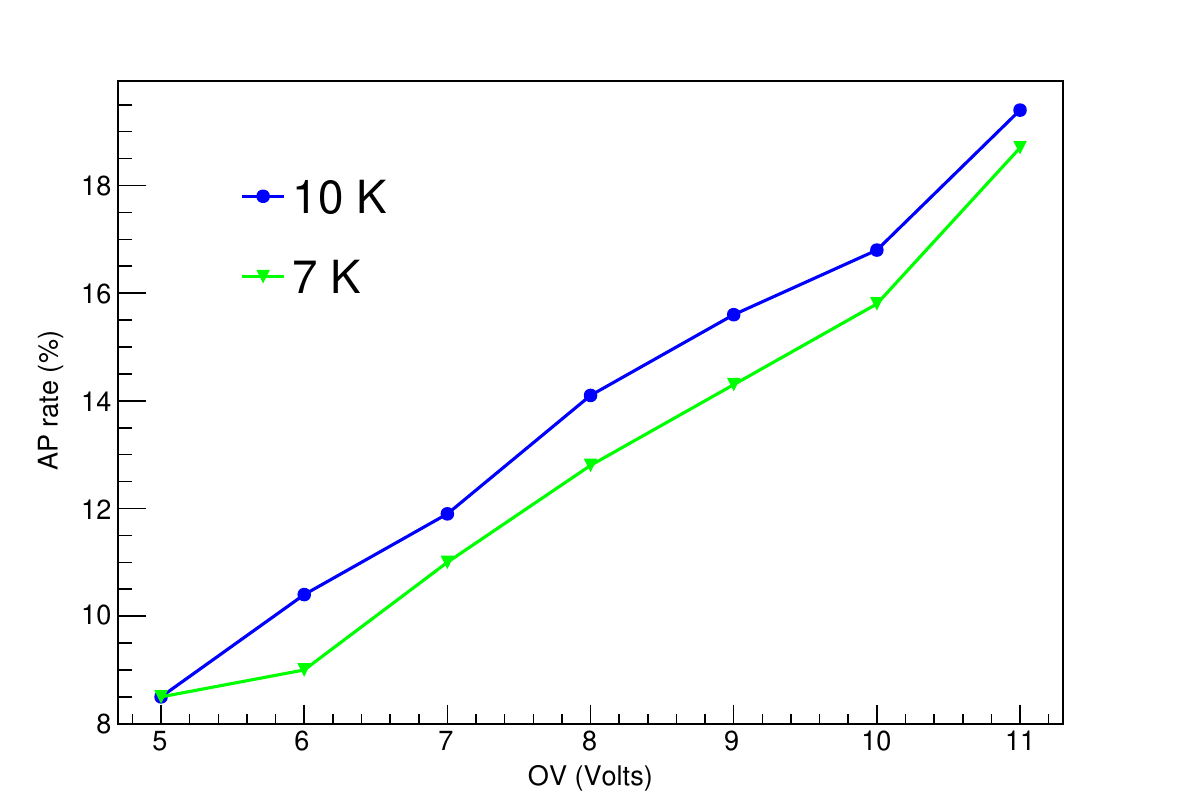}
	\caption{The combined AP results of an FBK SiPMs worked at 7 K and 10 K when biased with 9 V OV (35 V).}\label{combinedAP7K10K} 
\end{figure}

We re-measured the AP, CT, and PDE at 7 K again the same way as at 10 K. The combined results of AP at 7 K and 10 K are shown in Fig.~\ref{combinedAP7K10K}. The plot has the following features: (a) the higher the bias voltage, the greater the AP; (b) the AP at 10 K is slightly higher ($\sim$ 2 \%) than 7 K (for an OV greater than 6 V);  (c) for the SiPMs' possible working bias voltage, OV = 10 V, the measured AP is $\sim$ 16\% at 7 K and 10 K; and (d) with the 6 V OV, the AP we measured are $\sim$ 9\% and  10.5\% at 7 K and 10 K, respectively, which are consistent with reference~\cite{Acerbi17}. Although a complete understanding of the AP at cryogenic temperatures is still lacking, reference~\cite{Acerbi17} pointed out two conflicting processes that might help interpret the data. At a lower temperature, the SiPMs' quenching resistance would increase, which will (i) suppress the avalanche triggering probability in recharge due to a lower voltage applied on the single photon avalanche diode (SPAD) and (ii) have a longer time constant to enhance the chance of a carrier to be released to generate an AP event. However, no known mechanism exists to characterize the AP phenomena precisely at low temperatures. We are communicating with FBK people to better understand our AP results.

\begin{figure}[!t]	 
	\centering
    \includegraphics[width=0.8\textwidth]{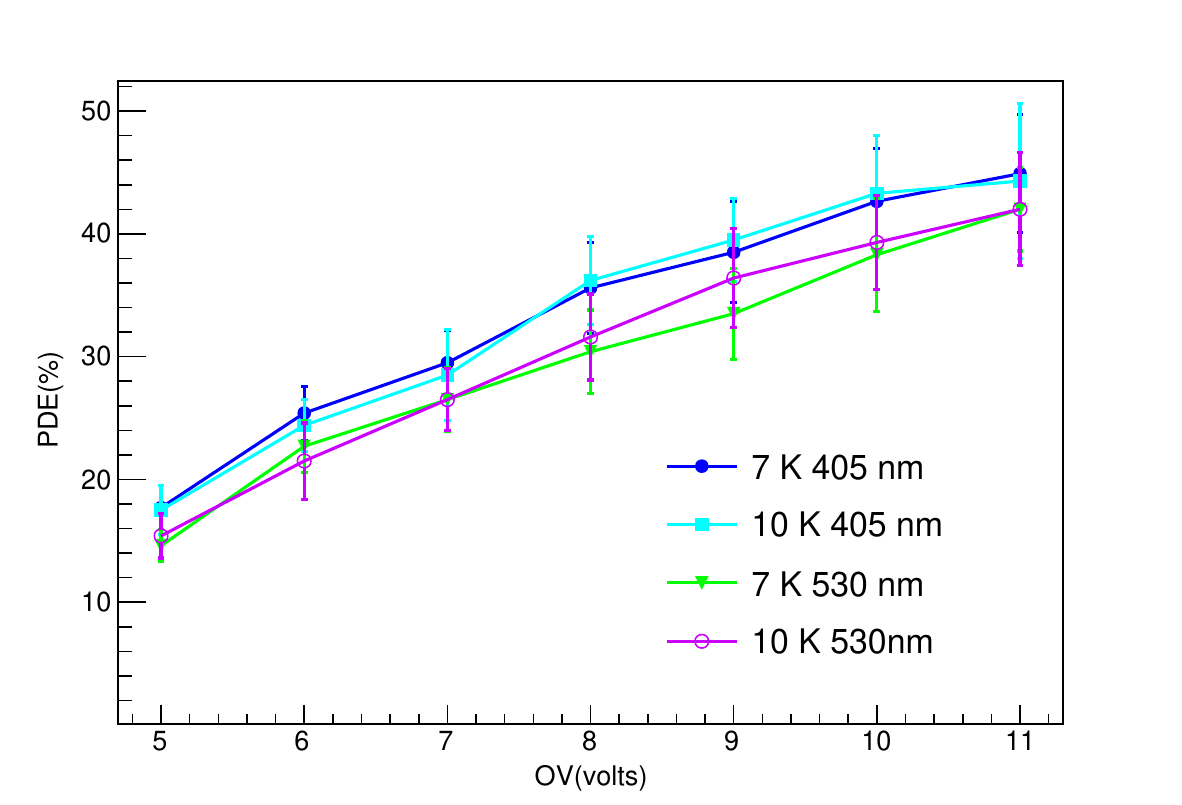}
	\caption{The combined PDE results of an FBK SiPMs worked at 7 K and 10 K.}\label{combinedPDE7K10K} 
\end{figure}

Regarding the PDE tests, Fig.~\ref{combinedPDE7K10K} does not show apparent temperature dependence at 7 K and 10 K. In addition, the plot shows that 405 nm light tends to generate a greater PDE than 530 nm, which is consistent with the tests on the similar type of SiPMs~\cite{Zappala16} at RT.

\section{Discussion and summary}\label{summaryDiscussion}

\subsection{Discussion}

\begin{figure}[!t]	 
	\centering
    \includegraphics[width=0.8\textwidth]{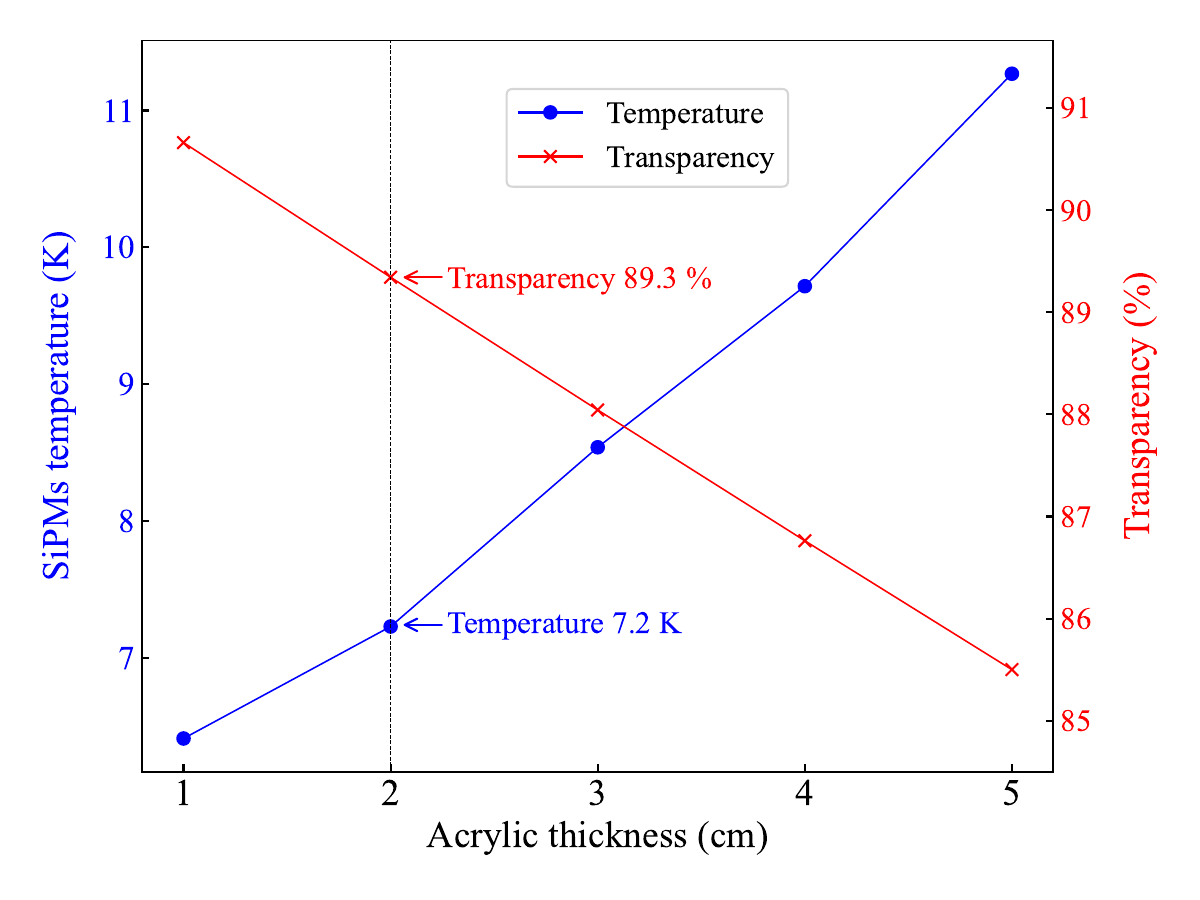}
	\caption{The COMSOL simulated temperature of the SiPMs and the calculated transparency of the acrylic layer on variant thickness. In the simulation, the liquid side of the acrylic layer was 4.5 K LHe, while the gas side was 1 atm helium gas; the SiPMs were mounted on the acrylic and surrounded by the gas. If the acrylic is 2 cm thick, the acrylic surface's temperature would be 7.2 K (and the SiPMs' temperature would be higher), as the left Y-axis shows; while the transparency would be 89.3\%, as the right Y-axis. We demonstrated in the manuscript that the FBK NUV-HD-Cryo SiPMs can be functional at 7.0 K.}\label{FigArylicThickTempTrans} 
\end{figure}

Although the lowest temperature the FBK NUV-HD-Cryo SiPMs have been verified to be functional so far is 7 K, instead of the liquid helium temperature of 4.5 K, due to the limitation of the cryocooler, we should be reassured about whether the SiPMs are eligible to work on an LHe TPC or not. The reason is that even if 7 K is the lowest temperature the FBK SiPMs can work, they can still be mounted on the TPC's top and bottom bases as photosensors provided the SiPMs and the LHe were isolated by a 2-cm thick acrylic. Fig~\ref{FigArylicThickTempTrans} indicates that a 2-cm thick acrylic would keep the SiPMs' temperature at 7.2 K while maintaining an 89.3\% transparency with the same assumption made by DarkSide-20k~\cite{DarkSide20k17}, a 1.5 cm acrylic layer would have a 90\% transmission for TPB-converted light.

\subsection{Summary}\label{SecDisSumSubSum}

We have measured the DCR, PDE, AP, and CT on the FBK NUV-HD-Cryo type SiPMs at 7 K and 10 K. The FBK SiPMs showed significantly different performance near LHe temperature than at RT. The DCR is only $\sim$ 0.01 Hz/mm$^2$ even with the bias voltage of OV = 12 V, which is equivalent to the Hamamatsu R11410 PMTs installed in the LZ TPC~\cite{LZTDR17}. The PDE reaches the desired 40\% when the bias voltage is OV 10 V near LHe temperature. AP and CT are both $\sim$ 10 - 15 \%, which are reasonably low for the photosensors working for rare events searching experiments. Combing all of the measurements together, we conclude that the FBK NUV-HD-Cryo SiPMs are suitable as photosensors to be equipped on LHe detectors, including but not limited to the ALETHEIA TPCs~\cite{ALETHEIA-EPJP-2023}.

\acknowledgments

We thank the FBK company in Trento, Italy, for sending us the SiPMs. In particular, we thank Dr. Alberto Gola, Dr. Alberto Mazzi, and Dr. Elena Moretti for SiPMs' delivery and other fruitful discussions. We especially thank Dr. Elena Moretti for reading the first version of our manuscript and addressing insightful comments. Dr. Xiangliang Liu and her team at NIM in China have kindly calibrated two photodiodes for us at 6 K on their facilities for free, which we greatly appreciate. In addition, we thank Dr. Wei Hu and Ms. Tangtang Qu at the TOFTEK company for their cooperation in developing readout electrics. We also appreciate Prof. Xiaoguang Wu, Prof. Baozhen Zhao, and Dr. Jinglong Wang at CIAE for their kindness in sharing their devices with us. Junhui Liao would also thank the support of the ``Yuanzhang'' funding of CIAE to launch the ALETHEIA program. This work has also been supported by NSFC (National Natural Science Foundation of China) under the contract of 12ED232612001001 and the ``Continuous-Support Basic Scientific Research Project'' in China.

\bibliography{SiPM2023}

\end{document}